\documentclass[12pt, preprint]{aastex}
\usepackage{amssymb}
\usepackage{amsmath}

\shorttitle{Bayesian/MCMC models for NGC 6503}
\shortauthors{Puglielli \& Widrow \& Courteau}

\begin{document}

\title{Dynamical Models for NGC 6503 using a Markov Chain Monte Carlo Technique}
\author{David Puglielli\altaffilmark{1}, Lawrence M. Widrow\altaffilmark{2} and St\'{e}phane Courteau\altaffilmark{3}
\affil{Department of Physics, Engineering Physics, and
Astronomy, Queen's University,
Kingston, ON, K7L 3N6, Canada}}

\altaffiltext{1}{dpuglielli@astro.queensu.ca}
\altaffiltext{2}{widrow@astro.queensu.ca}
\altaffiltext{3}{courteau@astro.queensu.ca}

\begin{abstract}
We use Bayesian statistics and Markov chain Monte Carlo (MCMC) techniques to construct dynamical models for the spiral galaxy NGC 6503. The constraints include surface brightness profiles which display a Freeman Type II structure; \ion{H}{1} and ionized gas rotation curves; the stellar rotation, which is nearly coincident with the ionized gas curve; and the line of sight stellar dispersion, which displays a $\sigma-$drop at the centre. The galaxy models consist of a S\'{e}rsic bulge, an exponential disc with an optional inner truncation and a cosmologically motivated dark halo. The Bayesian/MCMC technique yields the joint posterior probability distribution function for the input parameters, allowing constraints on model parameters such as the halo cusp strength, structural parameters for the disc and bulge, and mass-to-light ratios. We examine several interpretations of the data: the Type II surface brightness profile may be due to dust extinction, to an inner truncated disc or to a ring of bright stars; and we test separate fits to the gas and stellar rotation curves to determine if the gas traces the gravitational potential. We test each of these scenarios for bar stability, ruling out dust extinction. We also find that the gas likely does not trace the gravitational potential, since the predicted stellar rotation curve, which includes asymmetric drift, is then inconsistent with the observed stellar rotation curve. The disc is well fit by an inner-truncated profile, but the possibility of ring formation by a bar to reproduce the Type II profile is also a realistic model. We further find that the halo must have a cuspy profile with $\gamma \gtrsim 1$; the bulge has a lower $M/L$ than the disc, suggesting a star forming component in the centre of the galaxy; and the bulge, as expected for this late type galaxy, has a low S\'{e}rsic index with $n_b\sim1-2$, suggesting a formation history dominated by secular evolution.
\end{abstract}


\section{Introduction}

The joint analysis of surface brightness profiles and rotation curves has traditionally yielded rich insights into the physics of disc galaxies. Surface brightness profiles allow for bulge-disc decomposition, while extended rotation curves constrain mass models \citep{vanalbadaetal, palunaswilliams, debloketal, denarayetal}. However, mass models are subject to degeneracies due, in part, to uncertainties in the mass-to-light ($M/L$) ratios of the baryonic components (e.g., Maller et al. 2000, Dutton et al. 2005). Breaking these degeneracies requires more data, such as a line of sight (LOS) velocity dispersion profile or a colour profile. Additional data of this type allows for the creation of dynamical models \citep{rixetal, gebhardtetal, widrowperrettsuyu, baesdejonghe, thomasetal}. Dynamical models, in turn, provide the initial conditions for $N-$body simulations, which are needed to study the growth of bar and spiral instabilities. In addition, non-circular motions complicate the interpretation of gas rotation curves \citep{rheeetal, valenzuelaetal}.  Stellar rotation curve data may therefore help address these complications. A data set that includes surface brightness profiles, gas and stellar rotation curves and stellar velocity dispersions can provide constraints on fundamental galaxy parameters, along with information about stability to bar formation in the disc.

Such a data set exists for the isolated dwarf Sc galaxy NGC 6503. Gas and stellar rotation curves were measured by \citet{devaucouleurscaulet}, \citet{begeman87} and \citet{bottema89} (hereafter B89); B89 also measured the stellar velocity dispersion profile and surface brightness profiles in the $B-$ and $R-$bands. There are no strong asymmetries in galaxy structure (such as a bar), so it is a good candidate for analytic axisymmetric models.  However, further investigation reveals several peculiarities. The galaxy possesses a sharp drop in velocity dispersion near the centre (known as a `$\sigma-$drop'). The gas and stellar rotation curves are nearly coincident despite the standard asymmetric drift formula's prediction that stars should noticeably lag behind the gas. Finally, the surface brightness profile displays four distinct regions: a central peak, a flat region indicating a Freeman (1970) Type II profile, an inner exponential of scale length $\sim1$ kpc and a shallower outer exponential.

In this paper, we present several scenarios corresponding to differing interpretations of the data. For each scenario, we construct dynamical models that reproduce these features. We decompose the surface brightness using a S\'{e}rsic bulge and an inner-truncated light profile of the type used by \citet{kormendy77}. Inner truncated disc profiles may be due to dust extinction against an exponential disc \citep{macarthuretal}; they may be intrinsic to the density profile; or they may be due to a ring of bright stars which does not trace the mass. We explore these possibilities by testing models numerically under each scenario for bar formation. 
The coincidence of gas and stellar rotation curves also deserves attention as it suggests far less asymmetric drift than predicted. Thus, we wish to understand whether gas follows circular orbits that trace the gravitational potential, or whether it has its own asymmetric drift. We investigate this issue by testing two different ways of fitting the rotation curves: by fitting the stellar rotation alone and using the asymmetric drift to calculate the circular velocity; and by fitting the gas alone assuming it traces the circular velocity. 

Our aim is to construct dynamical models for each of these scenarios and test them via $N-$body simulations. We use the GalactICS model from Widrow et al. (2008, hereafter WPD) defined in terms of distribution functions for the disc, bulge and halo. The halo allows for a cusp or a core, and the bulge follows a S\'{e}rsic (1968) profile. The disc surface density profile is exponential with an optional inner truncation. 

The large number of parameters makes finding a fit to the data a challenging exercise. A number of techniques have been developed to determine best-fit parameters in such complex spaces -- notably, maximum likelihood techniques which involve minimizing the $\chi^2$ value. 
One promising approach which improves upon the maximum likelihood method employs Bayes' theorem and a Markov chain Monte Carlo (MCMC) technique to survey the relevant parameter space. Bayes' theorem provides a method of determining the posterior probability of a hypothesis based on both the likelihood function \emph{and} prior information about the input parameters; MCMC provides for a way to survey the parameter space. In conjunction, the two techniques provide the joint posterior probability distribution function (PDF) of the multidimensional parameter space; the marginal posterior PDF for any parameter (and hence its formal mean and error bars) are obtained by marginalizing over nuisance parameters. This technique yields realistic constraints on the model parameters, including (but not limited to) the halo cusp, the disc and bulge $M/L$ ratios, masses for each component, inner truncation parameters, and bulge S\'{e}rsic index. Similar techniques have been used, for example, to determine cosmological parameters \citep{tegmarketal, percival05, corlessking}.

In this work we build dynamical models for NGC 6503 using a Bayesian/MCMC method.  We fit several data sets for this galaxy \emph{simultaneously} in order to obtain as complete a picture of the galaxy as possible; however, the lack of a bar provides a further constraint. Therefore, we also obtain the stability parameters $X$ and $Q$ for our models. The $X-$parameter determines global stability to multi-armed modes \citep{toomre81} and the $Q-$parameter determines stability locally \citep{toomre64}. As a purely phenomenological matter, $Q$ can also determine global stability to bars \citep{athanassoulasellwood}. We test the stability of our models using $N-$body simulations.

Our approach allows us to investigate numerous other properties of the galaxy. The S\'{e}rsic index is indicative of bulge formation history \citep{courteauetal, kormendykennicutt}. We also explore the possibility that the bulge is related to luminous nuclear clusters found near the centres of many disc-dominated late-type galaxies, and derive $M/L$ ratios for the disc and bulge. Moreover, we can constrain the cuspiness of the halo. Cosmological simulations consistently demonstrate that halos are expected to have cuspy halos (e.g., Navarro et al. 1997; Moore et al. 1999), but 
there are considerable difficulties involved in inferring central halo densities from rotation curve measurements \citep{hayashietal, rheeetal, duttonetal, valenzuelaetal}. For example, gas rotation curve data near the centre of the galaxy may not trace the gravitational potential; using the stellar rotational velocity, where available, may be a better way to infer the cusp value \citep{pizzellaetal}. Further, noncircular gas motions at the centre (as may be caused by a triaxial halo) may skew the interpretation of the rotation curve, depending on the orientation of the disc \citep{simonetal}. With both gas and stellar rotation curves, NGC 6503 provides an excellent opportunity to investigate these issues. 

This paper is organized as follows. In \S2 we detail the observational properties of NGC 6503 and elaborate on the peculiarities outlined above. In \S3 we discuss our dynamical models, and in \S4 we describe our application of the Bayes/MCMC method. We discuss our primary stability results in \S5, in the process constraining the available physics, and we present additional results in \S6. Section 7 is the discussion, which includes an exposition of prior research on NGC 6503.

\section{NGC 6503: Observations and Issues}

NGC 6503 is an isolated dwarf Sc galaxy at a distance of 5.2 Mpc \citep{karachentsevsharina} and is inclined at $74\pm1^{\circ}$ \citep{begeman87}. It displays a small, bright central bulge, no signs of strong asymmetric central structure (e.g., no obvious bar) apart from some spirality, and it is mostly free of gas. In this section we review existing photometric and kinematic observations of the galaxy. 
Throughout this paper, we use a distance at 5.2 Mpc and therefore $1$~kpc~$= 40''$ at this distance.

\subsection{The surface brightness profile}

\begin{figure}
 \begin{center}
\includegraphics[angle=270,scale=0.8]{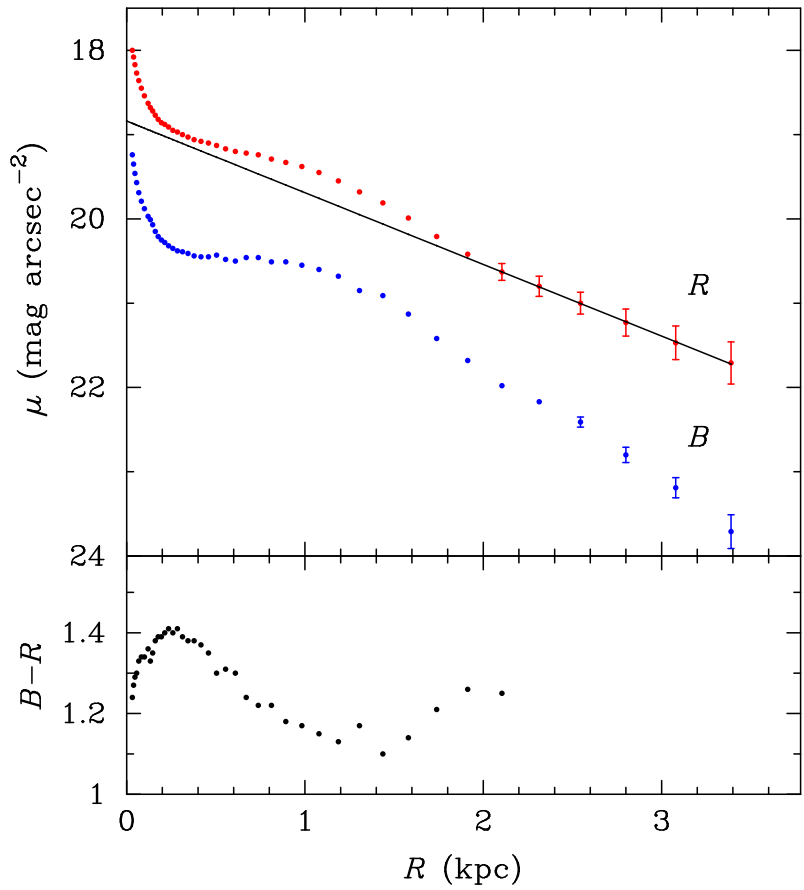}\end{center}
 \caption{Surface brightness and colour profile from Bottema (1989). The black curve shows a linear fit to the outer exponential with scale length $1.18\pm0.04$ kpc.}
 \label{fig:sbdata}
 \end{figure}

\begin{figure}
 \begin{center}
\includegraphics[angle=270,scale=0.8]{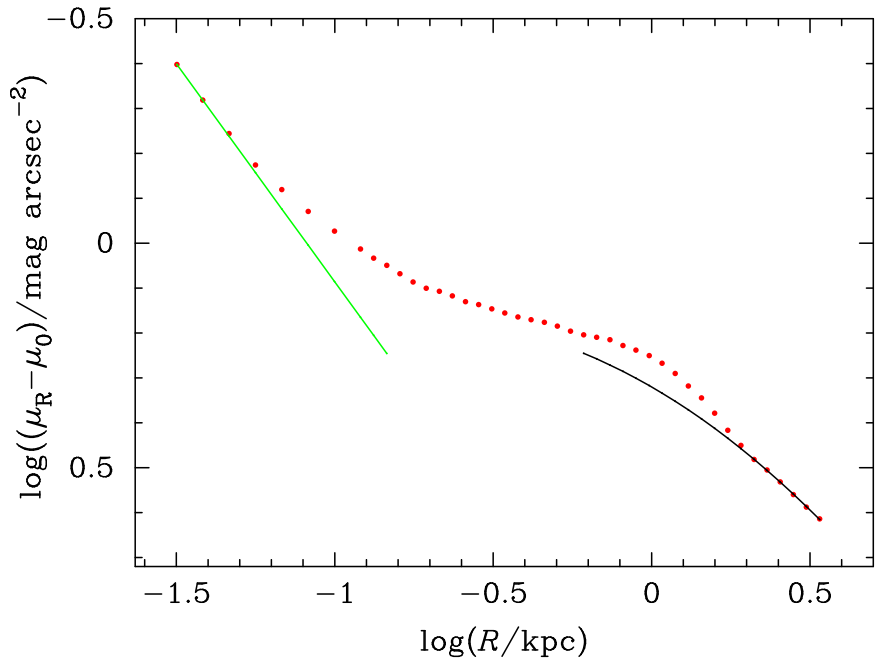}\end{center}
 \caption{A $\log\mu$-$\log R$ plot of the surface brightness. The green line is determined by the slope of the inner two points and has a slope of $\sim 1$. The black curve shows the fit found in Fig.~\ref{fig:sbdata}. The central magnitude $\mu_0$ is estimated from the slope of the inner points in Fig.~\ref{fig:sbdata} to be 17.6 mag arcsec$^{-2}$.}
 \label{fig:sbloglog}
 \end{figure}

Photometric imaging by B89 provides azimuthally averaged surface brightness (SB) profiles in the $B-$ and $R-$bands, both of which display a sharp rise within 50 pc and a Freeman (1970) Type II hump between 50 pc and $\sim 2$ kpc. The SB profile is exponential beyond that. The central flattening is accompanied by a slight reddening in the $B-R$ profile (Fig.~\ref{fig:sbdata}). 
A circumnuclear H$\alpha$ ring is observed at $R\sim1.2$ kpc, indicating a star-forming region \citep{knapenetal}. The outer exponential is remarkably straight in spite of the large error bars; a linear fit to those points yields a photometric scale length $R_d$ of $1.18\pm0.04$ kpc, which becomes about 1.3 kpc when corrected for inclination and scale height. 
Fig.~\ref{fig:sbloglog} shows the $\log\mu$-$\log R$ plot of the surface brightness. The inner few points are almost straight, approaching a line with a slope $\sim 1$, suggesting a S\'{e}rsic index of 1.

There are three possibilities for the origin of Type II profiles: first, they may be caused by dust extinction; second, they may reflect the formation of bright stars that do not trace the mass; and third, they may be intrinsic to the mass distribution of the disc. In NGC 6503, the reddening in the inner portion may support the dust extinction hypothesis (Bottema \& Gerritsen 1997, hereafter BG97). BG97 adopt an exponential disc and assume that the flattening in the surface brightness is due to dust (we return to this work in \S7.1). The possibility that dust extinction causes Type II profiles is also discussed by \citet{macarthuretal} for their sample of spiral galaxies with optical and IR imaging; these authors are unable to reliably distinguish between Type I and Type II profiles on the basis of dust extinction, however. 

The second possibility relates to bar formation. \citet{baggettbaggettanderson} and \citet{andersonbaggettbaggett} show that many galaxies with Type II profiles can be fit using Kormendy's (1977) inner truncated profile
\begin{equation}
I(R) = I_0 \exp[-(R/R_d + (R_h/R)^\alpha)]
\label{eq:kormendy}
\end{equation}
where $R_h$ is a turnover radius and $\alpha$ is a cutoff index. Most of Baggett et al.'s (1998) galaxies have a bar, but a significant minority do not. These authors also argue that such flattening is intrinsic to the galaxy rather than being caused by dust. It is known that bars can induce ring formation in galaxies, which may contribute to the Type II profile flattening, and that circumnuclear rings correlate strongly with barred galaxies \citep{knapen05}.  Moreover, simulations by \citet{foyleetal} suggest that the SB flattening could also be caused by the presence of a central bar, as the resulting reorganization of light can flatten the SB profile over the bar's extent. Furthermore, while NGC 6503 does not appear to have a bar, it is possible for bars to dissolve by gas inflow, which transfers angular momentum from the gas to the bar \citep{bournaudetal}. The ring currently seen in NGC 6503 may therefore be a relic of a now dissolved bar. In a scenario of ring formation by bar destruction, the outer exponential represents the `true' disc, and the Type II hump is mainly due to the formation of bright stars that do not trace the mass. The results in \citet{foyleetal} suggest that the Type II profile could be reproduced by the emergence of a bar in a disc of scale length 1.3 kpc. Therefore, it is incumbent to investigate the SB profile by fitting only the points exterior to 2 kpc. 

A third possibility is that Type II profiles are caused by a genuine mass deficit in the inner region, and therefore we also fit the full SB profile directly by assuming an inner truncated disc \citep{kormendy77}. This naturally produces a better fit to the hump than to the outer points; along with the first scenario, this scenario provides a direct comparison with the modelling of BG97, who did not account for the outer exponential. 

\subsection{Rotation curves}

\begin{figure}
\begin{center}
 \includegraphics[angle=270,scale=0.8]{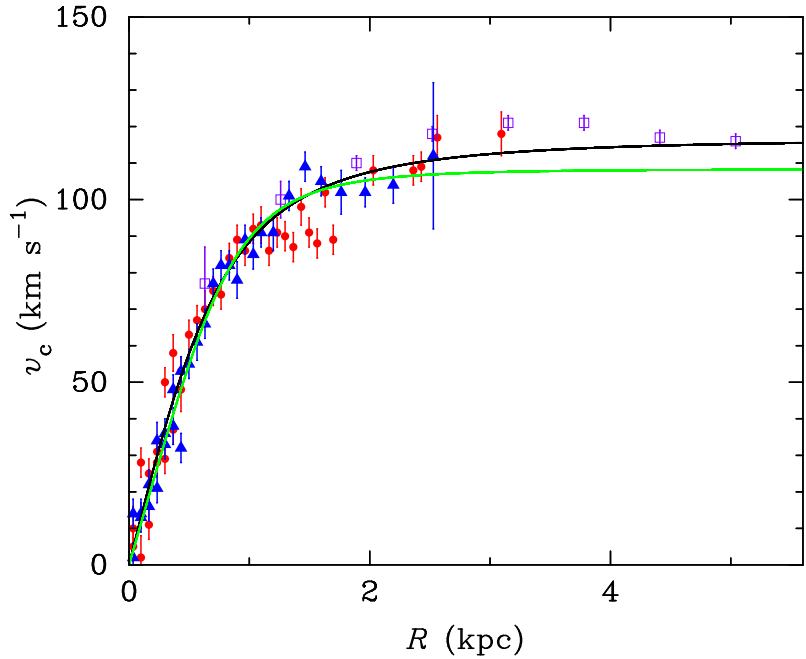}
\end{center}
 \caption{Rotation curve data for NGC 6503. The H$\beta$ data are filled red circles, the stellar rotation data are filled blue triangles (Bottema 1989) and the \ion{H}{1} data are hollow purple squares (Begeman 1987). The black curve shows the fit using Eq.~\ref{eq:courteau} to the gas data and the green curve shows the fit to the stellar data. The fitting parameters are: for the black curve, $v_0=117\pm1$ km~s$^{-1}$, $R_c=0.88\pm0.03$ kpc and $\zeta=2.1\pm0.1$; and for the green curve, $v_0=108\pm4$ km~s$^{-1}$, $R_c=0.95\pm0.04$ kpc and $\zeta=3.2\pm0.6$. The data have not been corrected for inclination.}
 \label{fig:Hrotationdata}
 \end{figure}

Several rotation curves for NGC 6503 have been measured. \ion{H}{1} observations by \citet{begeman87} reveal a remarkably flat rotation curve between 3 kpc and 20 kpc deviating by no more than $\sim4-5$\%, and a maximum circular velocity of $\sim 120$ km s$^{-1}$. \citet{greisenetal} find a similar \ion{H}{1} curve in addition to evidence for a thick and thin \ion{H}{1} disc in NGC 6503. The thick disc appears to rotate more slowly than the thin disc and the former does not extend beyond the optical radius. 
However, this distinction does not affect Begeman's measurements of the total \ion{H}{1} rotation curve. B89 maps the inner rotation curve with H$\beta$ and \ion{O}{3} emission as well as the stellar features, while \citet{devaucouleurscaulet} supply H$\alpha$ data. Where they overlap, all data sets are in agreement; the \ion{H}{1} observations appear to be slightly but systematically larger than the H$\beta$. The stellar rotation curve largely coincides with the H$\beta$ and \ion{O}{3} curves, with an average difference of only $\sim2$ km s$^{-1}$. At about 1.5 kpc, the stellar rotation exceeds the gas rotation, possibly owing to the effect of a spiral arm. For our models, we fit the \ion{H}{1}, H$\beta$ and stellar data sets. These are found in Fig.~\ref{fig:Hrotationdata}, along with fits based on the fitting formula \citep{courteau97}:
\begin{equation}
 v(r)=\frac{v_0}{(1+(R_c/R)^{\zeta})^{1/\zeta}}
\label{eq:courteau}
\end{equation}
where $v_0$ is an asymptotic velocity, $R_c$ is a projected length scale and $\zeta$ is a shape parameter governing the sharpness of the `turnover' to the asymptotic velocity. The fit suggests a projected asymptotic circular velocity of $117\pm1$ km s$^{-1}$ and shows how close the two rotation curves are.

The coincidence of the stellar and gas rotation curves must be accounted for. If the gas rotation traces the gravitational potential, the difference between the two rotation curves yields a measure of the asymmetric drift, $v_a$. The latter, defined as the difference between the circular velocity $v_c$ calculated from the potential and the streaming velocity $v_s$ of the stars, is given by
\begin{equation}
  v_a = \frac{\sigma_R^2}{2 v_c} \left[\frac{2R}{R_d}+\frac{1}{2}\left(\frac{R}{v_s}\frac{\partial v_s}{\partial R} - 1\right)\right]
 \label{eq:vasym}
 \end{equation}
for an exponential disc, where $R_d$ is the disc scale length and $\sigma_R$ is the radial velocity dispersion \citep{binneytremaine}. When applied to the observed H$\beta$ rotation curve, Eq.~\ref{eq:vasym} yields an asymmetric drift of order of $5-15$ km s$^{-1}$, depending on the adopted central radial dispersion.\footnote{B89 claims that the calculated asymmetric drift implied by the rotation curves is $\sim 3-5$ km s$^{-1}$, which is much lower than our calculated values. We have been unable to reproduce this number; every permutation of scalelength and dispersion that we tried in Eq.~\ref{eq:vasym} yields asymmetric drifts of at least $\sim$5 km s$^{-1}$. } 
The asymmetric drift is much larger than the observed difference between the gas and stellar curves; a possible cause of the discrepancy is that gas does not trace the gravitational potential. Traditionally, gas is assumed to travel on circular orbits, so that the H$\beta$/\ion{H}{1} rotation curves would trace the gravitational potential in this galaxy.  The assumption of circularity in gas rotation curves underlies much of the work on mass modelling in the literature. However, it is not possible for this assumption to strictly hold because gas dispersion is non-zero. There are several sources of this dispersion including turbulence caused by supernova feedback and the magnetorotational instability \citep{tamburroetal}. In addition, non-circular motions can impact the gas rotation curves of disc galaxies (e.g., Hayashi et al. 2004, Valenzuela et al. 2007), thereby reducing the observed velocity relative to the actual circular velocity. Being collisionless, our models cannot capture these effects; therefore, the hypothesis that the gas rotation traces the circular velocity can only be tested using the gravitational potential of our models.  However, the stellar asymmetric drift can easily be modeled to obtain a separate value for the circular velocity. The stellar asymmetric drift is itself subject to several caveats; it is, for example, highly sensitive to the orientation of the velocity ellipsoid. These issues are addressed in \S7.3.


\subsection{The LOS velocity dispersion}

\begin{figure}
\begin{center}
 \includegraphics[angle=270,scale=0.8]{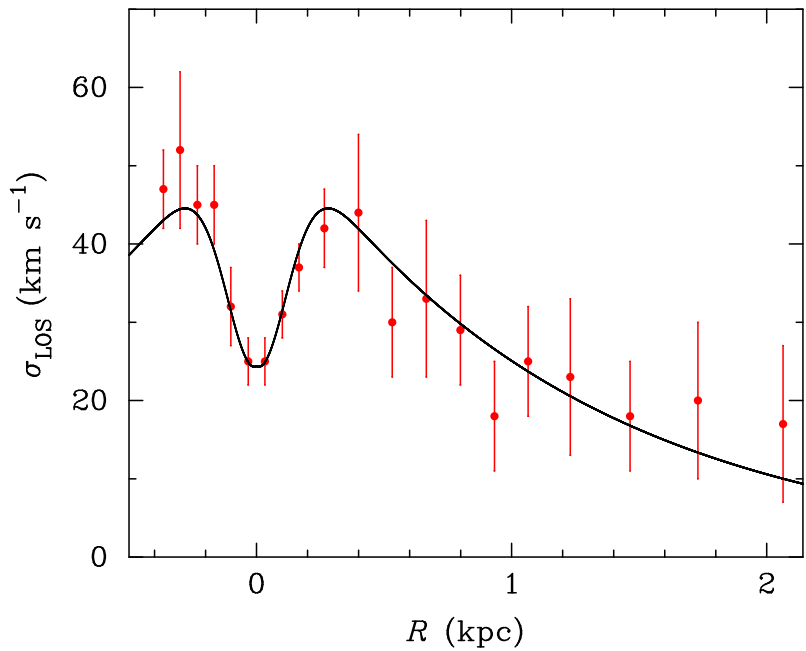}
\end{center} \caption{LOS stellar dispersion data from Bottema (1989) in red. The black curve shows an exponential + inverse Gaussian (Eq.~\ref{eq:gaussian}) fit to the data. The fitting parameters are given by $\sigma_0=59.5\pm8.4$ km s$^{-1}$, $R_{\sigma}=1.16\pm0.29$ kpc, $A=0.59\pm0.06$ and $B=0.12\pm0.02$ kpc. The data have not been corrected for inclination.}
 \label{fig:dispdata}
 \end{figure}

The LOS velocity dispersion profile for NGC 6503 from B89 is found in Fig.~\ref{fig:dispdata}. There is a ${\sigma}$-drop within 300 pc and an exponential decline beyond that range. We fit the data to an outer exponential with an inverse Gaussian in the centre:
\begin{equation}
 \sigma(R)=\sigma_0 \exp(-R/R_{\sigma})\left[1-A\exp(-R^2/(2B^2))\right]
 \label{eq:gaussian}
\end{equation}
where $A$, $B$, the extrapolated central dispersion $\sigma_0$ and the projected scale length $R_{\sigma}$ are the fitting parameters. The fit yields $R_{\sigma}=1.16\pm0.29$ kpc and $\sigma_0=59.5\pm8.4$ km s$^{-1}$, which will provide a point of comparison for our results in \S6. Note that we do not attempt to model the physical processes which may be responsible for the $\sigma-$drop, but we do determine structural parameters and effective $M/L$ ratios for the bulge to account for the $\sigma-$drop.

NGC 6503 was the first galaxy shown to have a $\sigma-$drop. A growing body of data suggests that $\sigma-$drops are found in numerous spiral galaxies \citep{marquezetal, comeronetal, delorenzocaceresetal}, although \citet{kovelaetal} argue that some $\sigma-$drops are observational artifacts. BG97 suggest that the $\sigma-$drop is due to a small central, possibly star forming component capable of continually cooling the centre. A physical explanation for $\sigma-$drops is proposed by \citet{wozniaketal} and \citet{champavertwozniak}, who argue that the presence of dynamically cold gas funnelling into the centre is responsible for forming dynamically cold stars that dominate the observed LOS kinematics. \citet{comeronetal}'s sample of $\sigma-$drop galaxies correlates with higher incidences of nuclear dust spirals, H$\alpha$ rings and Seyfert fraction, which would be consistent with a cold star-forming component in the centre due to gas inflow. Meanwhile, \citet{delorenzocaceresetal} suggest that inner nuclear bars may be responsible for at least some of the $\sigma-$drops found in the literature. 

\subsection{Four scenarios to test}

We test three different ways to fit the SB profile: (i) assuming an inner truncated disc; (ii) assuming an exponential disc with an inner truncated light profile to model simply the effect of dust extinction (a full dust model is outside the scope of this work); and (iii) fitting only the inner ($R\lesssim0.1$ kpc) and outer ($R>2$ kpc) points to account for the true disc being external to the Type II hump. Furthermore, we test two ways to model the rotation: (i) we fit the gas rotation by assuming that it equals the circular velocity; and (ii) we fit the stellar rotation by simulating the LOS observations of the inclined disc. Simulating observations along a line of sight is particularly important for stellar kinematic measurements because of the non-negligible scale height of the stellar disc, which leads to integration effects as the line of sight passes through regions of the disc that are not in the plane. This effect is less important for the gas disc, which is expected to have a small scale height. 

These tests are accounted for by four different scenarios. In scenario K (for Kormendy), we fit the full surface brightness with an inner truncated Kormendy disc, and we fit the stellar rotation curve. In KL (Kormendy light), we fit the surface brightness using an inner truncated light profile atop an exponential disc to account for dust extinction, and fit the stellar rotation as in scenario K. In scenario KG (Kormendy gas), we fit the surface brightness as in K but fit the gas rotation by assuming that the gas travels on circular orbits, ignoring the stellar rotation data. Finally, in scenario E (exponential), we fit the surface brightness excluding the points that constitute the Type II hump, and fit the stellar kinematic data as in K. Scenario E therefore alleviates the need for an inner truncated disc. In all cases, we fit the LOS velocity dispersion and the \ion{H}{1} data outside 3 kpc. These models and the fitting procedures are described in more detail in \S4 and 5; we first briefly discuss the GalactICS model and the MCMC technique.

\section{GalactICS Model}

Our goal is to construct a self-consistent equilibrium dynamical model for NGC 6503. Dynamical modelling yields the phase space distribution function (DF) for a system that precisely specifies its densities and velocities \emph{simultaneously}; for self-consistency and equilibrium, the model must satisfy both the Poisson and time-independent collisionless Boltzmann equations. The GalactICS model of Widrow et al. (2008, hereafter WPD) is designed to satisfy these criteria; it is derived from the \citet{kuijkendubinski} model. We briefly describe the GalactICS model here and refer the interested reader to \citet{kuijkendubinski} and WPD for details. 

\subsection{Model components}

\subsubsection{The bulge and halo}

The bulge is designed to reproduce the S\'{e}rsic profile \citep{sersic68, ciotti91} upon projection, for which the intensity is given by 
\begin{equation}
I(r) = I_0 \exp \left[ -b(\tilde{r}/r_b)^{1/n_b} \right]
\end{equation}
where $\tilde{r}$ is the projected radius, $I_0$ is the central intensity, $r_b$ is the half-light radius, $n_b$ is the S\'{e}rsic index of the bulge and $b$ is a parameter which depends on $n_b$. The expression is deprojected using an Abel integral equation to yield the intrinsic density distribution 
\begin{equation}
\rho_{\mathrm{bulge}}(r)=\rho_b \left(\frac{r}{r_b}\right)^{-p} \exp\left[-b\left(\frac{r}{r_b}\right)^{1/n}\right]
\end{equation}
where $r$ is the spherical radius and $p=1-0.6097/n+0.05563/n^2$ yields the S\'{e}rsic profile \citep{prugnielsimien}. We define a characteristic velocity scale $v_b$ for the bulge by
\begin{equation}
v_b =\left[4\pi nb^{n(p-2)}\Gamma(n(2-p))r_b^2\rho_b\right]^{1/2}
\end{equation}
and use this as an input parameter instead of $\rho_b$.

For the halo we adopt a cosmologically motivated profile (e.g., Navarro, Frenk \& White 1996, 1997; Moore et al. 1999; Diemand et al. 2005), given by
\begin{equation}
\rho = \frac{\rho_h}{(r/a_h)^{\gamma} (1+ r/a_h)^{3-\gamma}}
\end{equation}
where $\rho_h$ and $a_h$ are the halo scale density and radius, respectively, and $\gamma$ is the cusp value governing the shape of the halo; $\gamma=1$ yields a NFW profile, while $\gamma=0$ produces a core. A characteristic velocity scale $\sigma_h$ is introduced according to $\sigma_h^2 = 4\pi a_h^2 \rho_h$. For practical purposes, the density profile is truncated using an error function (WPD). We allow $\gamma$ to vary for maximum flexibility in fitting the observed data, since simulations of halo formation suggest that cusp values other than 1 may be appropriate \citep{mooreetal} and that halo profiles may not be universal \citep{jingsuto, navarroetal}.

\subsubsection{The disc and the Kormendy profile}

Surface brightness profiles in discs tend to decline exponentially with radius. Assuming light traces mass, this implies that the surface density is exponential:
\begin{equation}
 \Sigma(R)=\Sigma_0\exp(-R/R_d)
\end{equation}
where $R_d$ is the photometric scale length. Additionally, observations of edge-on galaxies suggest that disc galaxies have vertical sech$^2$ profiles with constant scale heights \citep{vanderkruitsearle}. Thus, the three dimensional density of the GalactICS disc is given by
\begin{equation}
\rho(R,z) = \rho_0 \exp(-R/R_d) \textrm{sech}^2(z/z_d)
\label{eq:expsech}
\end{equation}
where $z_d$ is the scale height. The DF corresponding to this density is given in \citet{kuijkendubinski}; it is an extension of Shu's (1969) planar DF into three dimensions.

However, as discussed in \S2, Type II profiles can also be modelled by Eq.~\ref{eq:kormendy}. If the light traces the mass, this suggests a surface density given by 
\begin{equation}
\Sigma(R) = \Sigma_0 \exp\left[-(R/R_d + (R_h/R)^\alpha)\right].
\label{eq:kormendy2}
\end{equation}
As before, the three dimensional density of the disc is given by
\begin{equation}
 \rho(R,z) = \rho_0 \exp\left[-(R/R_d-(R_h/R)^\alpha)\right] \textrm{sech}^2(z/z_d).
\label{eq:kormendy1}
\end{equation}
We modify the GalactICS model accordingly to generate this density. If Type II profiles are instead assumed to be due to dust extinction, Eq.~\ref{eq:expsech} holds for the disc density, and only the light follows an inner truncated profile (Eq.~\ref{eq:kormendy}).

\subsubsection{The dynamics}

The KD disc employs two integrals of motion: the energy $E$ and the angular momentum $L_z$, and an approximate third integral describing the energy in the vertical motions, $E_z$. By the Jeans theorem, any function of three isolating integrals of motion in a given potential will exactly solve the CBE \citep{binneytremaine}. The third integral allows for $\sigma_z \neq \sigma_R$; observations of the solar neighbourhood suggest that $\sigma_z/\sigma_R = 0.6$ \citep{wielen74}, which would be impossible if the DF were a function of only two integrals of motion. 

The GalactICS model decouples the vertical and radial dispersion of the disc. The vertical dispersion is given by the vertical potential gradient and scale height. The radial velocity dispersion is given by
\begin{equation}
 \sigma_R^2 = \sigma_0^2 \exp(-R/R_{\sigma})
\label{eq:radsigma}
\end{equation}
and the tangential dispersion is given by the epicycle equations, i.e. 
\begin{equation}
\sigma_{\phi} = \frac{\kappa}{2 \Omega} \sigma_{R}.
\end{equation}
Here, $\Omega=v/R$ is the angular rotation speed and the epicycle frequency $\kappa$ is given by $\kappa^2= \left[R(d\Omega^2/dR)+4\Omega^2\right]$. By design, the velocity ellipsoid in the GalactICS disc is cylindrically aligned.

Note that Eq.~\ref{eq:radsigma} includes a parameter to control the scale length of the velocity dispersion profile. Although observations by \citet{bottema93} suggest that $R_d = R_{\sigma}$, there is no clear theoretical reason why this should necessarily be the case. Here we treat $R_d$ and $R_{\sigma}$ independently to provide maximum flexibility in fitting all data sets.


\subsection{Summary of the GalactICS model}

The GalactICS model, modified to include inner truncation, has two further parameters for the disc; the turnover radius $R_h$ and the cutoff index $\alpha$. We complete the parameter input set with  $R-$band $M/L$ ratios for the bulge and disc. We assume that the $M/L$ ratios are constant with radius, and use the same value for the surface brightness and the stellar kinematic fits. Although the band pass over which the stellar observations were taken are not identical to the $R-$band, 
we do not expect the error introduced by this simplification to be significant.

Thus the parameters that we modify are: the five halo parameters; the exponential disc mass, scale height, scale length, and two Kormendy parameters; all bulge parameters; the central radial dispersion and dispersion profile scale length; and the disc and bulge $M/L$ ratios. In addition, each data set has an associated noise parameter that is allowed to vary (see \S4). We assume the halo to be nonrotating and truncate the disc at 5 kpc (well outside the measured data). The galaxy inclination $i$ is also fixed at 74$^\circ$. A list of the relevant parameters that are fit, as well as calculated output quantities, is presented in Table 1. The units used by GalactICS are such that $G=1$, but the masses have been converted back to $10^9M_{\odot}$ for simplicity. 

\section{Statistical Approach}

\subsection{Bayes' Theorem and MCMC}

The core idea of this study is to derive the joint posterior probability distribution function (PDF) of the galaxy parameters using Bayes' theorem and MCMC. Bayes' theorem provides a way to infer the posterior probability of a hypothesis, given some kind of evidence (such as observational data), using the prior probabilities of the evidence and model and a likelihood function (such as a $\chi^2$ function). The joint posterior probability distribution function (PDF) of the model parameters is written as $P(M|D)$, where $M$ is the set of model input parameters and $D$ is the set of observational data. Bayes' theorem may be written as
\begin{equation}
 P(M|D)=\frac{P(M)P(D|M)}{P(D)}
\label{eq:bayes}
\end{equation}
where $P(M)$ is the prior probability of the input parameters, $P(D|M)$ is the likelihood of the data given a specific set of model parameters and $P(D)$ is the prior probability of the data and functions as a normalization. 

MCMC, in turn, provides a method of surveying the parameter space that rapidly converges to the posterior probability distribution of the input parameters. Our MCMC technique explores the parameter space by way of the Metropolis-Hastings algorithm. Essentially, a random walk is constructed in parameter space by sampling a new set of model parameters $M^*$ at each step of the chain, and $P(M^*|D)$ is calculated using the likelihood function and the priors. The ratio $s=P(M^*|D)/P(M|D)$ is then calculated. If $s>1$, $M^*$ is accepted; otherwise, $M^*$ is accepted or rejected with a uniform random probability $\varrho$. Over the course of many such iterations, the chain populates the posterior PDFs, in the process supplying the PDFs for each parameter by marginalizing over the full PDF. 

Initially, the chain quickly travels to regions of `good' fits from any set of initial parameters, and eventually settles into the best fit region, a process known as `burn-in'. It can be proven that the resulting density of points in parameter space samples the real probability density distribution of the parameter space \citep{gregory}. We aim towards an acceptance rate of 23\% in our chains \citep{robertsgelmangilks}. A detailed description of the technique may be found in \citet{gregory}.

The benefits of MCMC are numerous: it fully samples the PDFs for all fitted parameters, meaning that formal errors and error contours are easily determined; it is easily extended to include more parameters if desired; it tends to move to a region of high likelihood fairly rapidly; other quantities, such as the stability parameters $Q$ and $X$, are easily calculated from the PDFs; and there is some evidence that it provides more realistic error constraints than maximum likelihood methods \citep{kellyetal}.

The primary disadvantages are twofold. First, there is no guarantee of uniqueness to the final solutions -- multimodal PDFs are possible in any given parameter space. However, a judicious choice of the step size and proper tweaking to obtain the correct acceptance rate mitigates this problem. Second, although MCMC chains tend to converge quickly, it is not so simple to verify that the converged chains have fully populated the joint posterior PDF. Moreover, 
the chain may `spread around' the parameter space, reflecting either the real structure of the joint PDF or the fact that the chain has not run long enough to fully populate the PDF. Formally ensuring convergence may require prohibitively long computing time. However, in practice we are able to obtain good constraints for most parameters in a reasonable time. This does not preclude the possibility that a MCMC chain passed through multiple PDF peaks after burn-in, but there would be no reason to prefer any one such peak over another.

\subsection{The likelihood function and the priors}

The likelihood function is given by 
\begin{equation}
\mathcal{L} = (2\pi)^{-N/2} \prod_{i=1}^{N} \sigma_i^{-1}\exp{\left[\frac{(d_i - m_i)^2}{2\sigma_i^2}\right]}
\end{equation}
where $N$ is the total number of data points, $d_i$ is the experimental data value, $m_i$ is the mean value predicted by the model and $\sigma_i$ is the error, scale-modified as described below. This expression is applied to each data set, and the total likelihood is given by the product of the individual likelihoods.

For the data values $d_i$, we use the H$\beta$ curve, the LOS dispersion, stellar rotation curve and the $R-$band SB profile from B89 and the extended \ion{H}{1} curve from \citet{begeman87}. We ran several different MCMC chains (described in the next section) making use of different data sets.

The model values $m_i$ are calculated as follows. For the gas, we assume that the observed H$\beta$/\ion{H}{1} rotation curves follow circular orbits so that the model gas rotation curve can be calculated directly from the potential. For the stellar rotation curve and dispersion profile, we sample the disc and bulge DFs along the line of sight at each data point from B89, appropriately weighted by the $M/L$ ratios. Finally, for the surface brightness we sample the DFs along elliptical annuli around the inclined galaxy centre at an axis ratio determined by the inclination angle (i.e., $q=\cos i=0.28$) to determine the elliptically averaged density and thus the surface brightness in mag arcsec$^{-2}$. We emphasize that in no cases have we corrected the observed profiles for inclination; all inclination effects are accounted for by rotating the \emph{model} galaxy and sampling along the line of sight.


For each dataset used in a particular MCMC run, a single noise parameter is implemented. This consists of an extra term added in quadrature to each error bar: $\sigma_i^2 = \sigma_{0i}^2 + f^2$, where $\sigma_{0i}$ is the original experimental error and $f$ is the noise parameter, taken as constant for all points in a given data set (so there is a total of four scale factors). This approach allows us to account for effects that cannot be captured by our models - for example, that of spiral structure on the light profile when the original models are axisymmetric. Noise parameters use a Jeffreys prior:
\begin{equation}
p(f) = \frac{1}{f \ln(f_{\mathrm{max}}/f_{\mathrm{min}})}.
\end{equation}
(Gregory 2005) where $f_{\mathrm{min}}$ and $f_{\mathrm{max}}$ refer to the minimum and maximum values of the noise parameters adopted for the MCMC run. Otherwise, $P(M)$ is uniform for all parameters. 

\section{Results}

\subsection{Dust extinction and inner truncation}

As discussed in \S2, the flattening in the SB profile may reflect the intrinsic mass distribution in the disc, or it may be due to dust extinction. We conduct two MCMC chains to examine these possibilities. For the first run, corresponding to scenario K, the mass distribution of the disc is given by Eq.~\ref{eq:kormendy1}, so that the disc is intrinsically truncated. We fit the full $R-$band SB profile from B89. In the second run, corresponding to scenario KL, the mass distribution of the disc is given by Eq.~\ref{eq:expsech} so that the disc is purely exponential and the light follows an inner truncated profile. For both runs, we fit the full stellar rotation curve, the \ion{H}{1} curve outside 3 kpc and the LOS stellar dispersion profile; we do not fit the gas rotation internal to 3 kpc here.

\begin{figure}
\begin{center}
 \includegraphics[angle=270,scale=0.6]{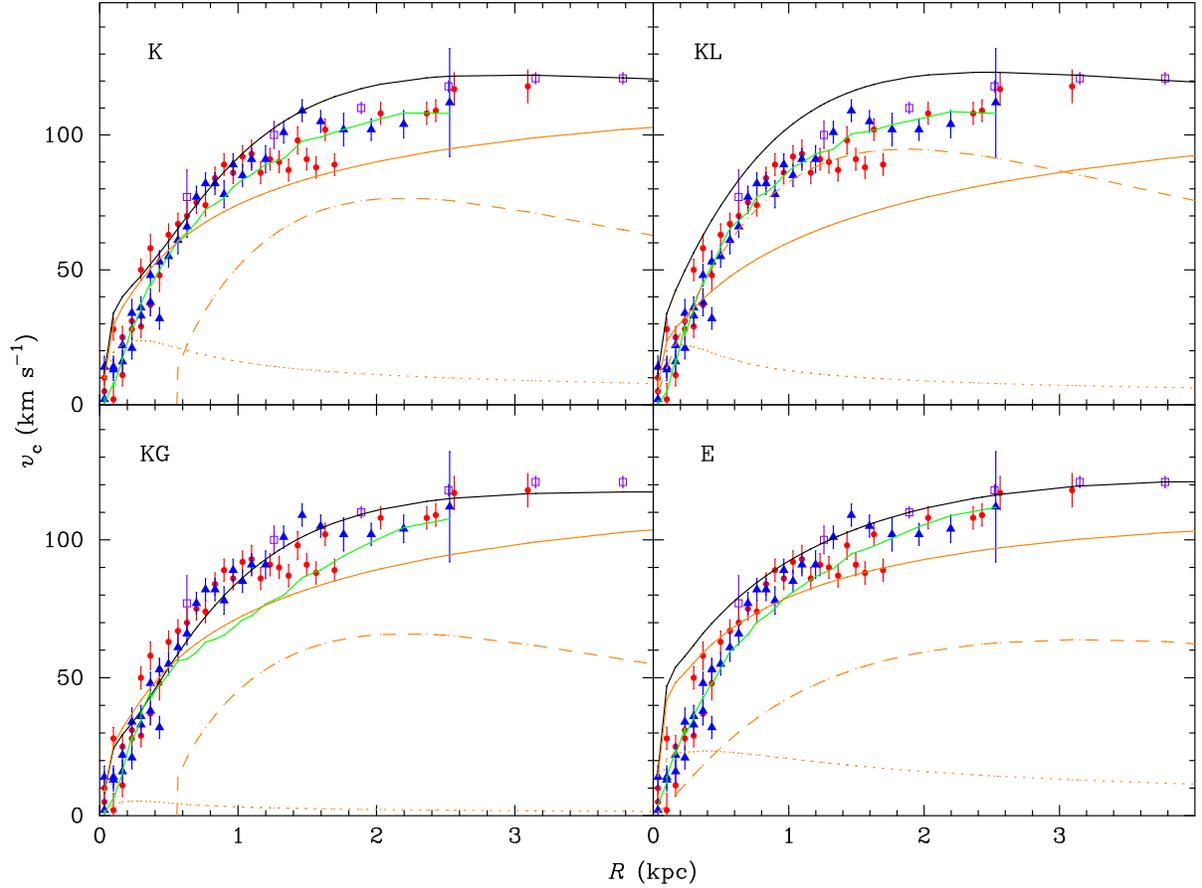}\end{center}
 \caption{Graphs of the rotation curves for representative models of each scenario. Red identifies the H$\beta$ points, purple identifies the \ion{H}{1} data and blue represents the stellar rotation points as in Fig.~\ref{fig:Hrotationdata}. Green is the stellar model circular velocity curve while black represents the circular velocity curve. The orange curves show the rotation curve breakdown by component: halo (solid), disc (dashed) and bulge (dotted). The scenario K and KG rotation curves display a kink in the black curve at $\sim 0.2$ kpc, owing to the inner truncation.}
 \label{fig:kormendy-model}
 \end{figure}

\begin{figure}
\begin{center}
 \includegraphics[angle=270,scale=0.8]{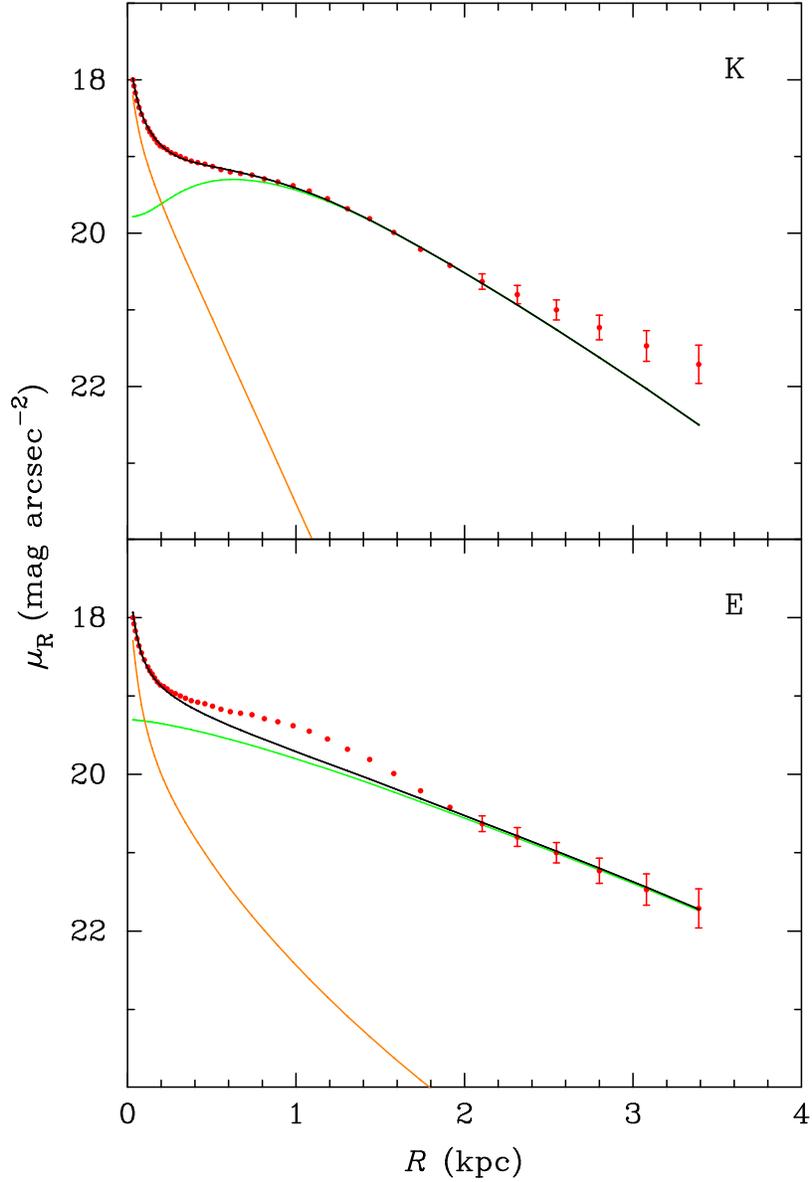}\end{center}
 \caption{Graphs of the SB profile for scenarios K and E. Orange is the bulge contribution to the surface brightness, green is the disc contribution, black is the total and red identifies the observed $R-$band surface brightness. All models from scenarios K, KL and KG show fits similar to the top panel.}
 \label{fig:kormendylight-model}
 \end{figure}

\begin{figure}
\begin{center}
 \includegraphics[angle=270,scale=0.8]{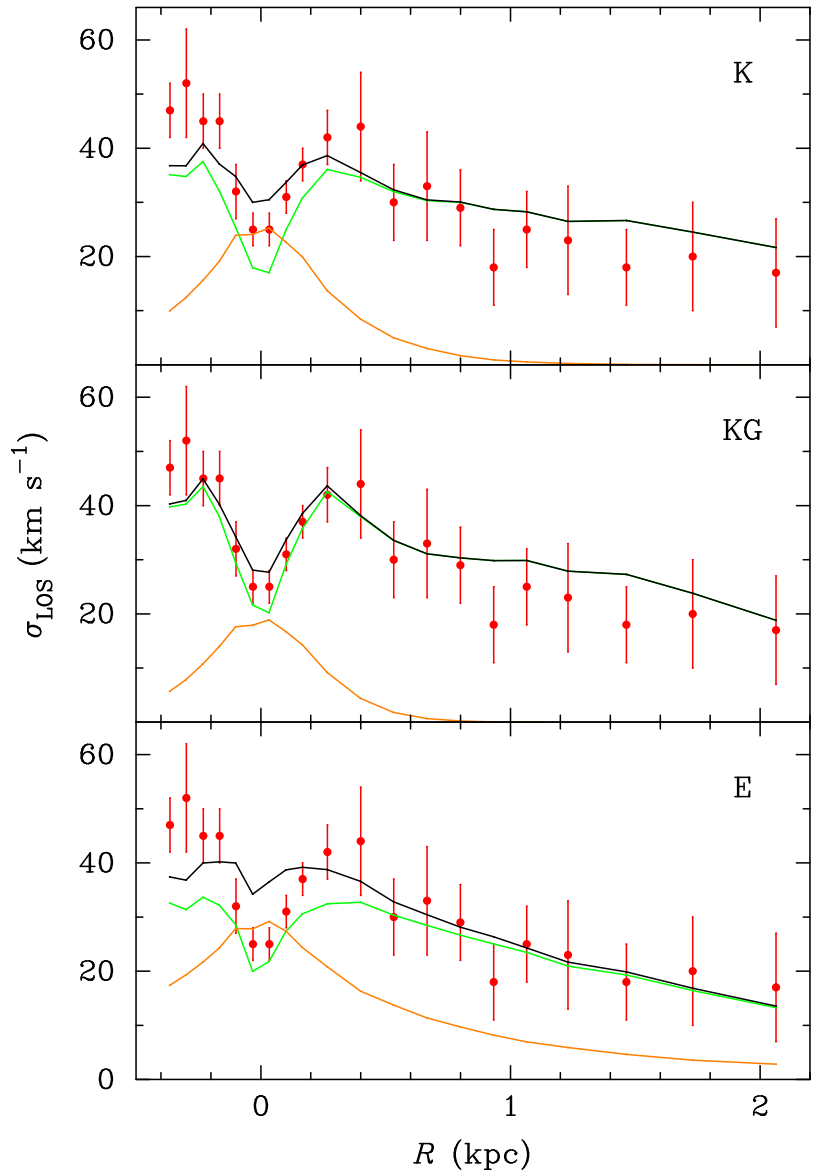}\end{center}
 \caption{Graphs of the LOS dispersion for scenarios K, KG, and E. The model dispersion is in black and the observed dispersion is in red. The bulge contribution to the dispersion is in orange and the disc contribution is in green. Fits for scenario KL are similar to those for scenario KG.}
 \label{fig:kormendygas-model}
 \end{figure}

Fits for representative examples of each scenario are found in Figs.~\ref{fig:kormendy-model}$-$\ref{fig:kormendygas-model}. In Fig.~\ref{fig:kormendy-model} we can see that the models reproduce the stellar rotation well. As expected, the circular velocity for these models is always greater than the H$\beta$ data. The rotation curve breakdown is also seen here, showing that the inner rotation curve for scenario KL is strongly disc-dominated. In addition, the surface brightness fit in Fig.~\ref{fig:kormendylight-model} is excellent for both runs within 2 kpc. The outer exponential is virtually ignored by the fit owing to the larger error bars. The dispersion profile fits in Fig.~\ref{fig:kormendygas-model} are also good. Scenario K displays a slightly inferior dispersion profile fit, owing to the lower $\sigma_0$ for these models, but the $\sigma-$drop is present. The outer region fit appears to overshoot the points slightly, owing to the large $R_{\sigma}$ for these runs. We have verified that decreasing $R_{\sigma}$ by 50\% produces better fits to the dispersion without affecting the remaining fits, suggesting that the cause of the poorer dispersion fit in the outer region is a local PDF peak. The $\chi^2$ values for scenario KL are approximately equal to those for scenario KL; therefore, there is no reason to reject this scenario on the basis of the fit quality. However, analysis of the PDFs and the time series trace of the $\chi^2$ value reveals that this run actually passes through two separate minima which correlate with two distinct regions of parameter space, both of which produce nearly identical fits.

\begin{figure}
\begin{center}
 \includegraphics[angle=270,scale=0.8]{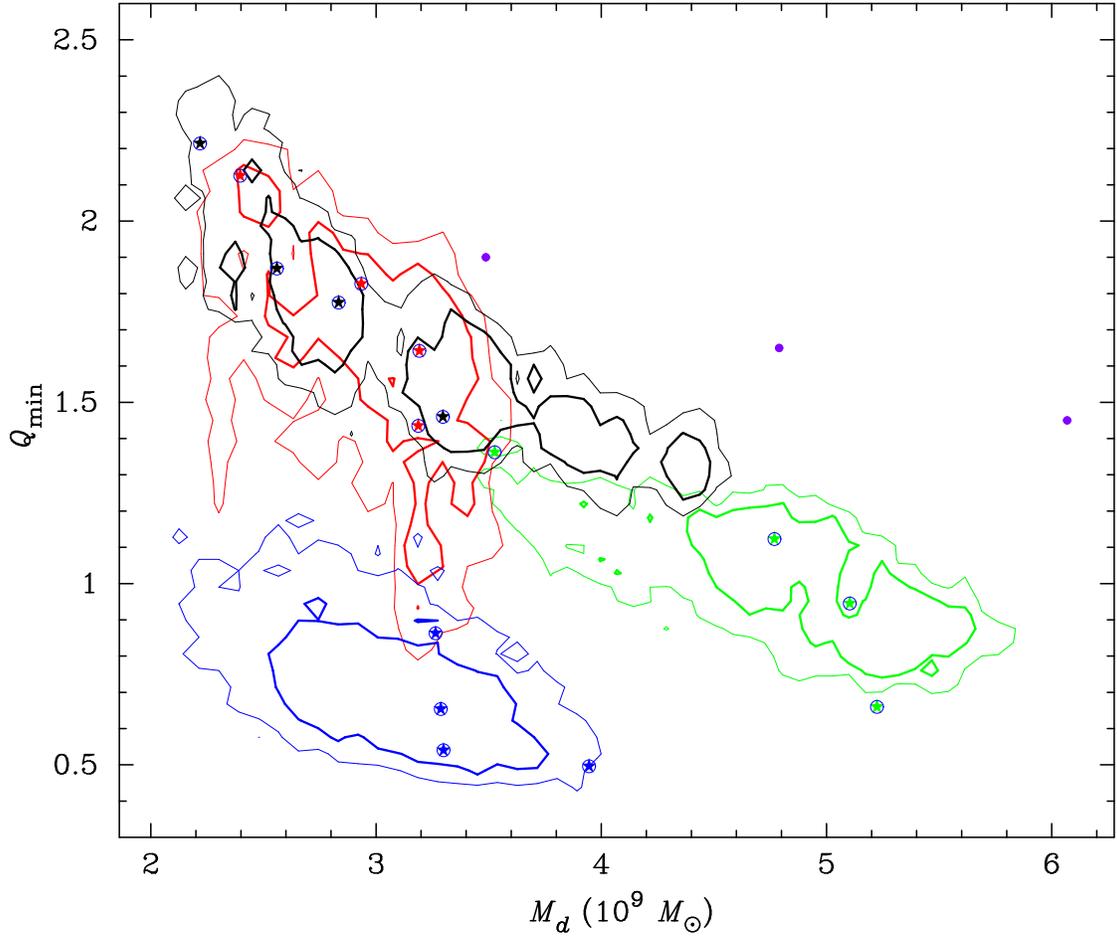}\end{center}
 \caption{The 2-D PDF  of the disc mass $M_d$ versus the minimum $Q$, $Q_{\mathrm{min}}$, of the disc. Black refers to scenario K; green refers to scenario KL; red refers to scenario KG; blue refers to scenario E. The thick lines enclose the 1$\sigma$ confidence region, and the thin lines enclose the 2$\sigma$ confidence region. The blue-outlined stars correspond to the parameters of the models chosen to test stability, and the purple circles correspond to the points used by BG97 (\S7.1).}
 \label{fig:kormendyall-discmass-Q}
 \end{figure}

Fig.~\ref{fig:kormendyall-discmass-Q} shows the two-dimensional PDF for $M_d$\footnote{Note that the input $M_e$ parameter is the exponential disc mass. It is distinguished from the true disc mass $M_d$ because the exponential disc must be truncated in the outer region, and because the inner truncation in scenarios K and KG removes additional mass from the inner region.} and the minimum disc $Q$, $Q_{\mathrm{min}}$, usually attained at around $1-1.5$ disc scale lengths. Scenario K is in black and scenario KL is in green. 
For scenario K, we find that the PDF spans a large range in $M_d$ and $Q_{\mathrm{min}}$, forming a relatively narrow trough from $Q_{\mathrm{min}}\sim2.3$ at $M_d\sim 2.1\times10^9 M_{\odot}$ in the upper left to $Q_{\mathrm{min}}\sim1.3$ at $M_d\sim 4.4\times10^9 M_{\odot}$ in the lower right. Not surprisingly, $M_d$ for scenario KL is much larger, since there is no hole in the mass distribution. Since the radial dispersion $\sigma_R$ and epicycle frequency $\kappa$ are constrained by the data, the increased surface density yields lower values for $Q_{\mathrm{min}}$. 
For scenario KL, 
the best fit region extends below the $Q_{\mathrm{min}}=1$ line, which suggests that some of the best fit models are locally unstable. 

\begin{figure}
\begin{center}
 \includegraphics[angle=270,scale=0.8]{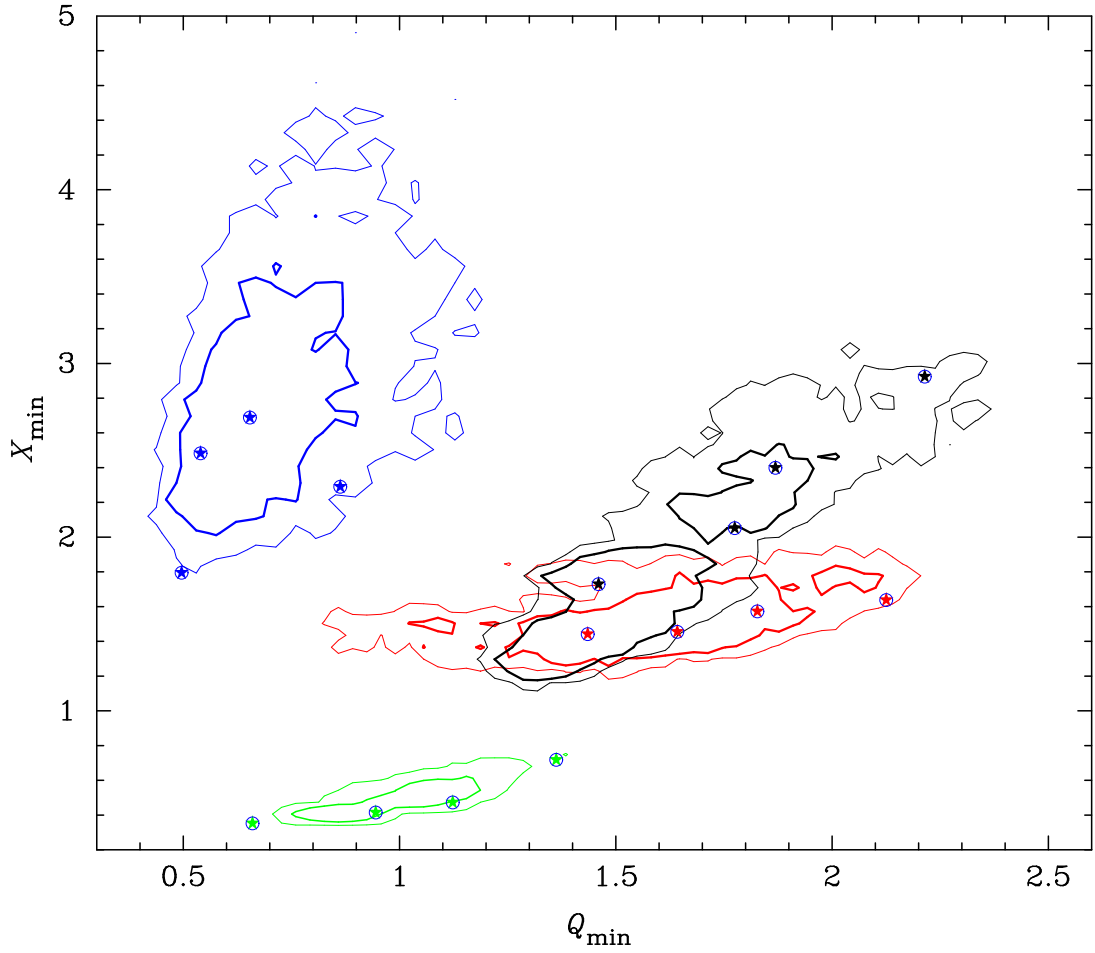}\end{center}
 \caption{The 2-D PDF of the minimum $X$, $X_{\mathrm{min}}$, versus minimum $Q$, $Q_{\mathrm{min}}$, of the disc. Colours and symbols are as in Fig.~\ref{fig:kormendyall-discmass-Q}.}
 \label{fig:kormendyall-Q-X}
 \end{figure}

The PDF in $Q_{\mathrm{min}}$ and $X_{\mathrm{min}}$ is found in Fig.~\ref{fig:kormendyall-Q-X}. For scenario K, this plot shows that $X_{\mathrm{min}}$ lies between 1 and 3, and appears to correlate weakly with $Q_{\mathrm{min}}$, in the sense that models with high $Q_{\mathrm{min}}$ also have high $X_{\mathrm{min}}$. Fig.~\ref{fig:kormendyall-Q-X} also shows that \emph{all} models from scenario KL have $X_{\mathrm{min}}<1$, suggesting very strong instability to global modes. 

To discern whether or not the models for scenario KL are realistic, we must determine whether or not galaxies with $Q<1$ are possible. Unstable galaxies generate spiral structure that heat stars as they propagate outwards but continual cooling could occur if gas infall onto the surface of the galaxy were substantial enough. Such gas would contribute to star formation. \citet{fuchs99} conducted a maximum disc analysis for NGC 6503 suggesting that $Q<1$ implies a star formation rate of 40 $M_{\odot}$ yr$^{-1}$; this contradicts the rate of 1.5 $M_{\odot}$ yr$^{-1}$ obtained from the observed H$\alpha$ flux \citep{kennicuttetal}. Therefore, it is unlikely that this cooling method is important in NGC 6503. In the absence of star formation, a disc in which $Q<1$ may fragment as small scale perturbations grow due to the Toomre instability. The likely evolution for such models is to eventually recombine into a single stable disc, but not before the stellar dispersions have increased dramatically.

\begin{figure}
\begin{center}
 \includegraphics[angle=270,scale=0.6]{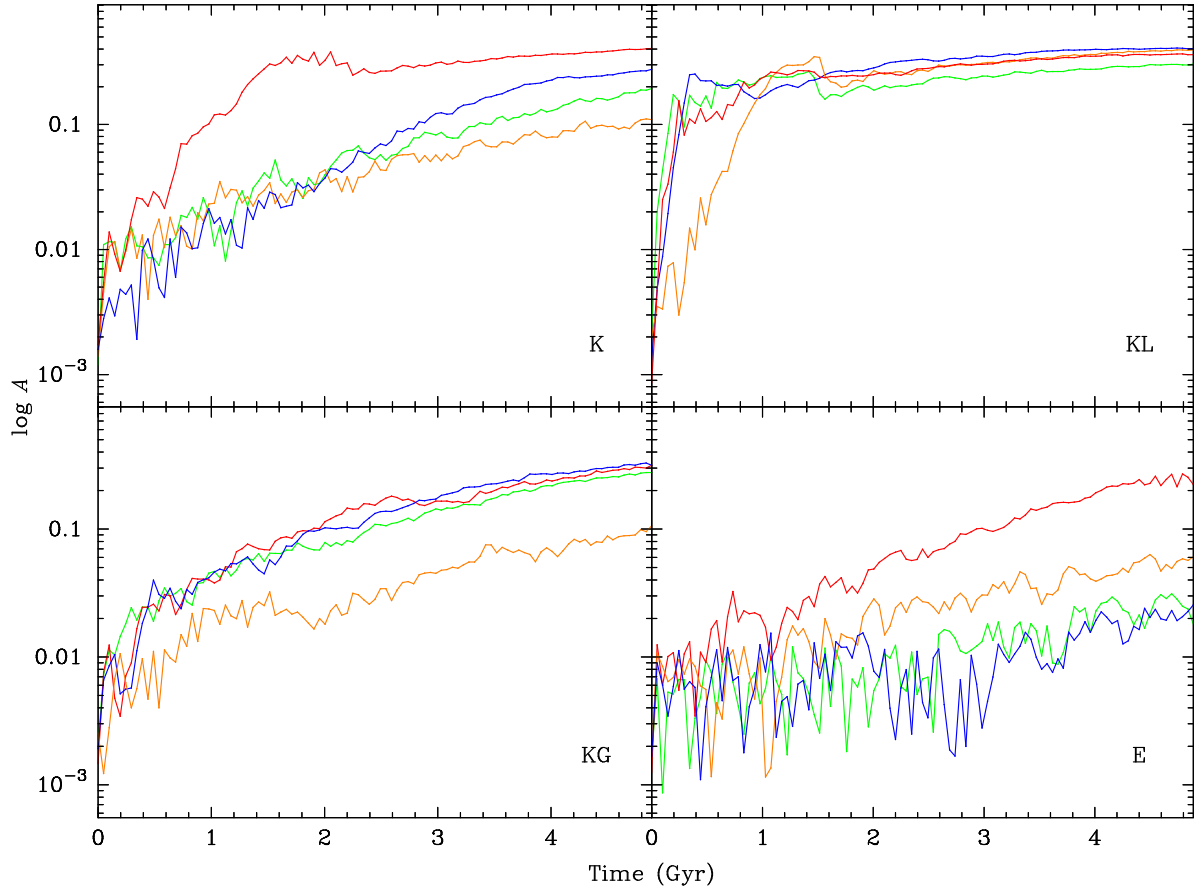}\end{center}
 \caption{The Fourier amplitude ln $A$ of several models obtained from each scenario. The colour pattern orange-green-blue-red corresponds to lowest to highest $Q_{\mathrm{min}}$ of the models identified by stars on Figs.~\ref{fig:kormendyall-discmass-Q} and \ref{fig:kormendyall-Q-X}.}
 \label{fig:kormendyall-bar}
 \end{figure}

We select several models from the PDF for both runs (identified by stars on Fig.~\ref{fig:kormendyall-discmass-Q}) and evolve them forward in time to examine their stability. To do this, we employ Dehnen's (2000) $N$-body algorithm implemented by \citet{stiff03}, a fast multipole method that produces nearly $O(N)$ scaling. All models are evolved for 5 Gyr with a time-step of $0.5$ Myr and a softening parameter of 25 pc. We use 500K particles for the disc, 50K for the bulge and 1M for the halo. The bar evolution is quantified using the magnitude of the second Fourier mode $A$, which measures the strength of two-armed asymmetries (e.g., Shen \& Sellwood 2004):
\begin{equation}
A= \left|\sum^{N}_{j=1}\frac{\exp(2i\theta_j)}{N}\right|
\end{equation}
where $\theta_j$ is the azimuthal coordinate of the $j$th particle in the disc. The bar evolution is found in Fig.~\ref{fig:kormendyall-bar}.

The results show the regions of the $M_d-Q_{\mathrm{min}}$ plane that are most susceptible to bar formation. For scenario K, we find that much of the region is at least mildly bar unstable, with the lower right region more strongly unstable (note that the red curves in Fig.~\ref{fig:kormendyall-bar} correspond to the discs with lowest $Q_{\mathrm{min}}$ in each run) and the upper left only showing a mild bar instability. The growth of most bars in this scenario is gradual, with bars only starting to become discernable after $\sim 2$ Gyr. Additionally, the central hole gradually vanishes in the bar unstable models as stars stream into the centre during the process of bar formation.

By contrast, the models for scenario KL are far more susceptible to global instabilities than for scenario K, as the growth of the bar mode is rapid and nearly instantaneous. We therefore rule out models from scenario KL as being unable to properly model the galaxy; this conclusion is a setback for the hypothesis that the Type II SB profile is due to dust extinction.



\subsection{Testing the rotation curve fits}

We next test fits to the gas and stellar rotation curves via a third MCMC chain corresponding to scenario KG. We fit the H$\beta$/\ion{H}{1} rotation curves by assuming that they trace the circular velocity and ignoring the stellar rotation data. The dispersion and surface brightness are fit as in scenario K. The stars will rotate more slowly because of asymmetric drift (Eq.~\ref{eq:vasym}), an effect that is incorporated into the GalactICS disc.

Examining the rotation curve fits in Fig.~\ref{fig:kormendy-model}, it is apparent that scenario K cannot reproduce the H$\beta$/\ion{H}{1} data and scenario KG cannot fit the stellar rotation data. Thus, it is likely that scenario K represents a more realistic model for NGC 6503. Note that both runs match the \ion{H}{1} data beyond $\sim3$ kpc, since there is no asymmetric drift at those radii. The other fits, shown by the surface brightness in Fig.~\ref{fig:kormendylight-model} and the LOS dispersion in Fig.~\ref{fig:kormendygas-model}, are more than satisfactory.

Fig.~\ref{fig:kormendyall-discmass-Q} shows that scenario KG displays considerable overlap with scenario K except in the higher mass range. The plot reveals that the best fitting models for this scenario span a large range in $Q_{\mathrm{min}}$, and here some models (albeit very few) extend down to below $Q=1$. However, Fig.~\ref{fig:kormendyall-Q-X} shows that all models for scenario KG are confined to $X_{\mathrm{min}}<2$, unlike for scenario K where many models reside between $X_{\mathrm{min}}=2$ and $X_{\mathrm{min}}=3$. Also unlike scenario K, $Q_{\mathrm{min}}$ does not appear to be significantly correlated with $X_{\mathrm{min}}$, and unlike scenario KL, in neither scenario K nor KG do we see $X_{\mathrm{min}}$ drop below 1.

As before, we selected several models from the PDF for scenario KG and evolved them forward in time. The results are found in the bottom panel of Fig.~\ref{fig:kormendyall-bar}. As for scenario K, most of the available parameter space is bar unstable, with the low mass-high $Q$ part of the plot susceptible only to mild bar formation. Thus, it is not possible to distinguish between runs K and KG on the basis of bar stability alone. Referring back to the rotation curve fits, we see that no models for scenario KG can reproduce the stellar rotation curve because the asymmetric drift is too large. Therefore, we reject models from scenario KG as being inconsistent with the data; however, see \S7.3 for a discussion of the issues involved in calculating the asymmetric drift.

\subsection{Testing the outer exponential fit}

\begin{figure}
 \begin{center}
\includegraphics[angle=270,scale=0.8]{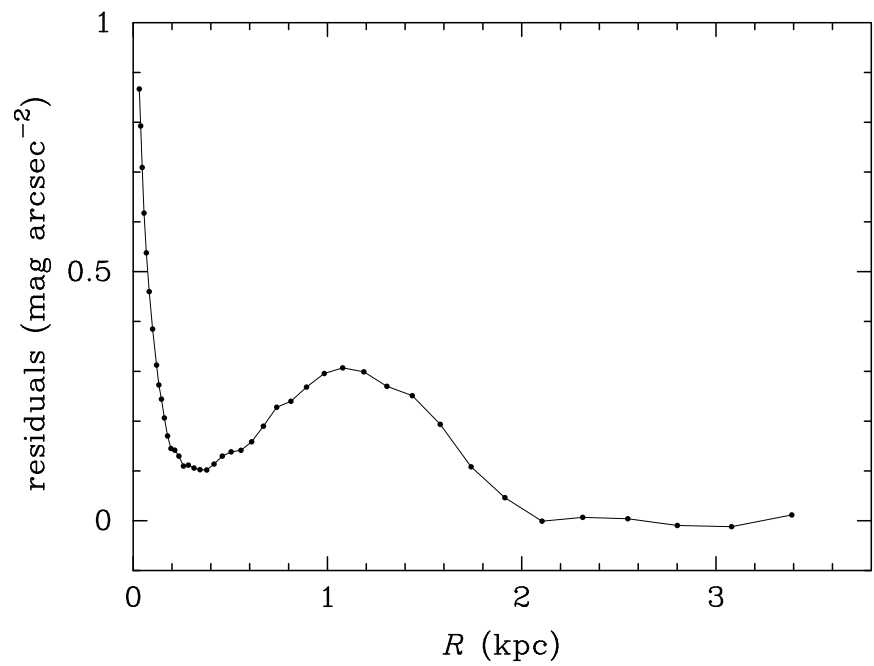}\end{center}
 \caption{Residuals in the $R-$band surface brightness for the exponential disc fit shown in Figure~\ref{fig:sbdata}. }
 \label{fig:sbresiduals}
 \end{figure}

Finally, we test the hypothesis that the underlying disc scale length is given by the outer ($>2$ kpc) portion of the SB profile by only fitting the interior points and the external points of the SB profile with a MCMC chain corresponding to scenario E. The outer cut is made at 2 kpc; $R_d$ is subtly affected by the inclination and differences in scale height, and, moreover, the very tight correlation between $M_d$ and the disc $M/L$ makes it difficult to obtain reasonable acceptance rates with MCMC. Therefore, we manually tested combinations of $R_d$ and $M/L$ to produce the best fit; we adopt a value of $R_d=1.3$ kpc and fix the disc $M/L$ to $0.53M_d$. To avoid biasing the resulting value for $z_d$, we do not include the external disc in our $\chi^2$ calculations. The inner cut is trickier; we have found that small changes to the number of inner points fit may lead to large changes in the resulting bulge parameters. To rectify this issue, we plot the residuals of the surface brightness from a regression line fit of $R_d=1.18$ kpc (Fig.~\ref{fig:sbresiduals}) and only fit the points that lie above the maximum of the Type II hump. At these points, the surface brightness is most assuredly bulge dominated. The stellar rotation is fit as in runs K and KL, along with the \ion{H}{1} data in the outer part, and the velocity dispersion is fit as in the other three runs. 

For this run, we assume that the evolution of a disc would generate a ring of star formation that could reproduce the Type II hump. The process of generating this profile would alter the rotation and dispersions, however, and we caution that the kinematics that result may not fit the data as well as the direct fits presented here using MCMC. Accounting for this would require ad hoc selections of what points to fit, or excluding one or more data sets entirely; for example, excluding the stellar rotation and only fitting the \ion{H}{1} data outside 3 kpc. Such a process is probably no more reliable than simply fitting the full data sets, evolving a test model from the PDFs and then determining how well the rotation curve and dispersion profile fit the observed data after evolution. Because we are assuming that a bar is supposed to form in this scenario, unstable models would not rule out this interpretation. However, models with $Q<1$ must be tested separately.

As in scenario KL, the stellar rotation curve is well fit while the circular velocity is larger than the H$\beta$ curve (Fig.~\ref{fig:kormendy-model}). The surface brightness fit is very good over the ranges we wish to fit; the inner eight points and outer exponential are well reproduced (Fig.~\ref{fig:kormendylight-model}). The velocity dispersion profile looks different from that for the previous scenarios since the bulge contributes significantly to the profile. Because of this, the $\sigma-$drop is not properly reproduced (S\'{e}rsic profiles do possess a slight central decline in the velocity dispersion -- see Ciotti 1991 -- but this feature cannot reproduce the observed $\sigma-$drop). Thus, we posit that either an additional nuclear component is responsible for the $\sigma-$drop, or that bar formation in this scenario would generate gas inflow that could reproduce the $\sigma-$drop. The disc contribution to the velocity dispersion is comparatively muted, resulting in a lower overall disc dispersion.

From Fig.~\ref{fig:kormendyall-discmass-Q}, we see that scenario E behaves differently than the other scenarios. These models have lower $M_d$ and $Q_{\mathrm{min}}$ but Fig.~\ref{fig:kormendyall-Q-X} shows that these models have high $X_{\mathrm{min}}$. Additionally, $Q_{\mathrm{min}}$ is reached at larger radii than in the other scenarios (typically, at $\sim2-2.5R_d$), and they have more massive bulges (\S6.4), which may also help inhibit the bar instability. 

\begin{figure}
 \begin{center}
\includegraphics[angle=270,scale=0.8]{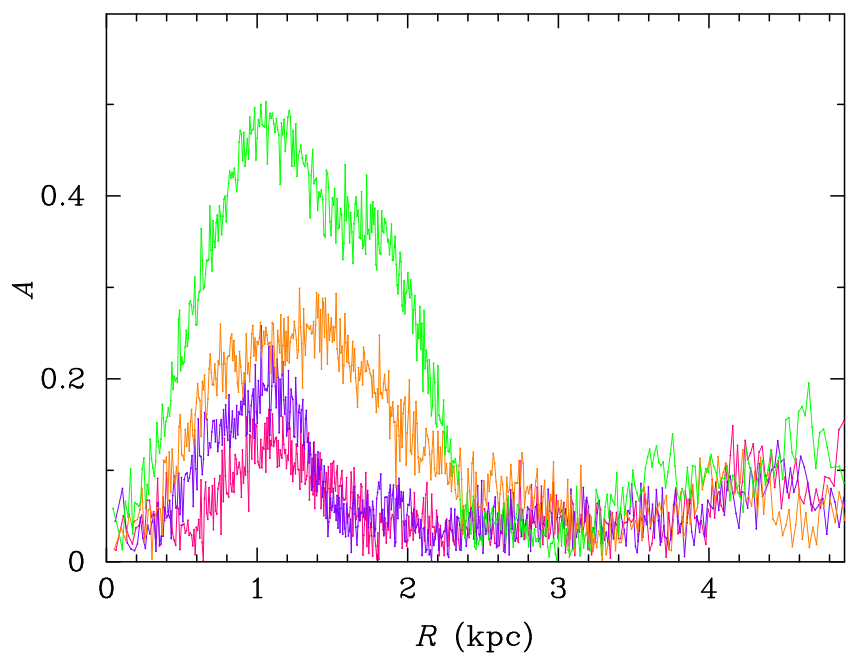}\end{center}
 \caption{Bar strength $A$ as a function of radius at different times for the lowest $X$ model from scenario E. Pink is the bar strength at 1.5 Gyr, purple at 2 Gyr, orange at 3 Gyr, and green at 4 Gyr.}
 \label{fig:exp-barlength}
 \end{figure}

As before, we selected several models and tested them for bar stability; the results are shown in the bottom right panel of Fig.~\ref{fig:kormendyall-bar}. While these models quickly form spiral structure, bar formation is gradual at the low $X_{\mathrm{min}}$ end of the best fit region and there is no evidence of bar formation at larger $X_{\mathrm{min}}$, despite the very low $Q_{\mathrm{min}}$. Because these simulations do not include gas, it is not possible to simulate possible bar destruction by gas inflow, but we can examine the length of the bar that forms. In Fig.~\ref{fig:exp-barlength} we plot the bar strength $A$ as a function of radius at different times for the lowest$-X$ model. This plot provides an estimate of the bar length of $\sim1.5$ kpc at $1.5-2$ Gyr, which is consistent with the location of the Type II hump. We conclude that scenario E is a realistic scenario for the formation of the bar.

\section{Further results}

In this section we delineate numerous other results that can be gleaned from the model. Because scenarios KG and KL are ruled out, we focus mainly on the results for scenarios K and E.

\subsection{The disc}

\begin{figure}
\begin{center}
 \includegraphics[angle=270,scale=0.8]{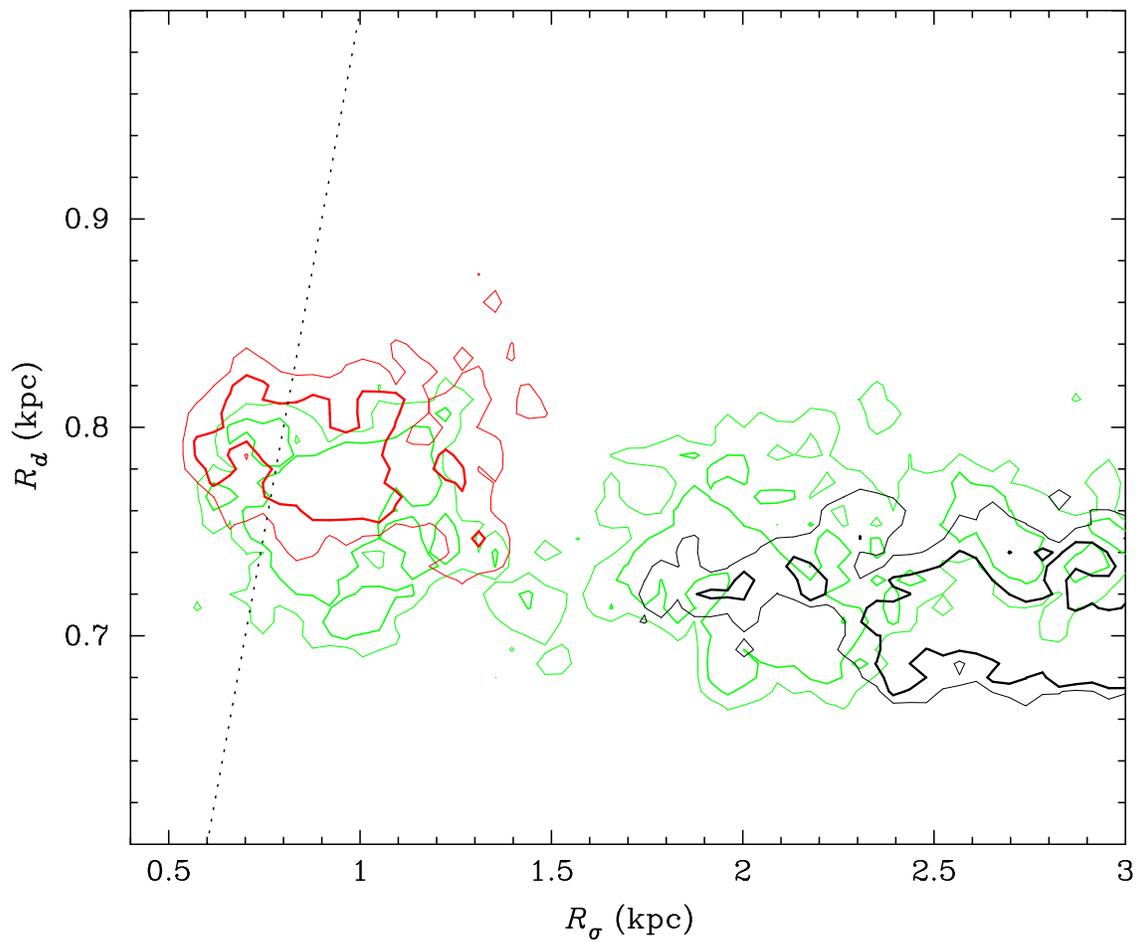}\end{center}
 \caption{The 2-D PDF of the photometric scale length $R_d$ versus the dispersion scale length $R_\sigma$. Colours are as in Fig.~\ref{fig:kormendyall-discmass-Q}. The dotted line identifies the $R_d=R_{\sigma}$ line.}
 \label{fig:kormendyall-disclength-sigmalength}
 \end{figure}

We begin by comparing the photometric and dispersion scale lengths. We find from scenario K that the dispersion scale length is significantly larger than the photometric scale length, as seen in Fig.~\ref{fig:kormendyall-disclength-sigmalength}. 
The mean $R_{\sigma}=2.6$ kpc is also considerably larger than the best fit $R_{\sigma}$ of 1.16 kpc found in \S2, while the photometric scale length is slightly less than the `fit by eye' value of 1 kpc from B89. Note that the dispersion scale length is poorly constrained owing to the size of the error bars in the velocity dispersion data, and that a wide range of $R_{\sigma}$ may produce good fits, as Fig.~\ref{fig:kormendyall-disclength-sigmalength} clearly shows. The large $R_{\sigma}$ results from the low extrapolated central dispersion $\sigma_0$ of only $37$ km s$^{-1}$, much less than found by direct fit in \S2. 
In scenario E, by contrast, the dispersion scale length of $R_{\sigma}=0.75$ kpc is much lower than the adopted photometric scale length of $R_d=1.3$ kpc. Here $\sigma_0$ is also very low, but the bulge dispersion is high enough to contribute much more to the dispersion profile than in the other scenarios, as is apparent from Fig.~\ref{fig:kormendygas-model}. Both scenarios suggest that $\sigma_0$ and $R_{\sigma}$ are not well constrained because of the large error bars in the LOS data.

\begin{figure}
\begin{center}
 \includegraphics[angle=270,scale=0.8]{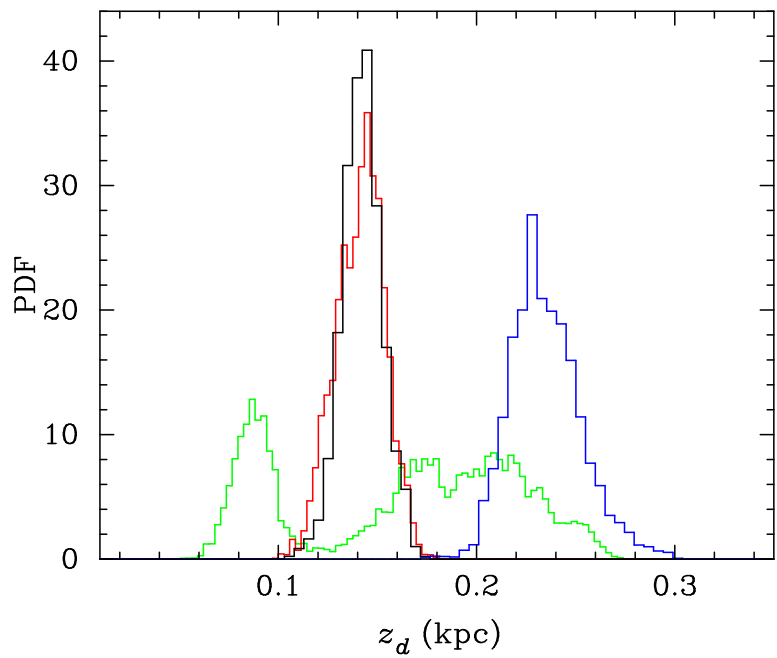}\end{center}
 \caption{The PDF for the disc scale height $z_d$. Black refers to scenario K; green refers to scenario KL; red refers to scenario KG; blue refers to scenario E.}
 \label{fig:kormendyall-heightPDF}
 \end{figure}

For the scale height $z_d$,  we find a value of $0.14\pm0.01$ kpc in scenario K and $0.24\pm0.02$ kpc in scenario E (Fig.~\ref{fig:kormendyall-heightPDF}). 
Both of these values are about one-fifth of the photometric scale length in scenarios K and E respectively, and are consistent with observations of the scale height in edge-on disc galaxies \citep{kregeletal}.

\begin{figure}
\begin{center}
 \includegraphics[angle=270,scale=0.8]{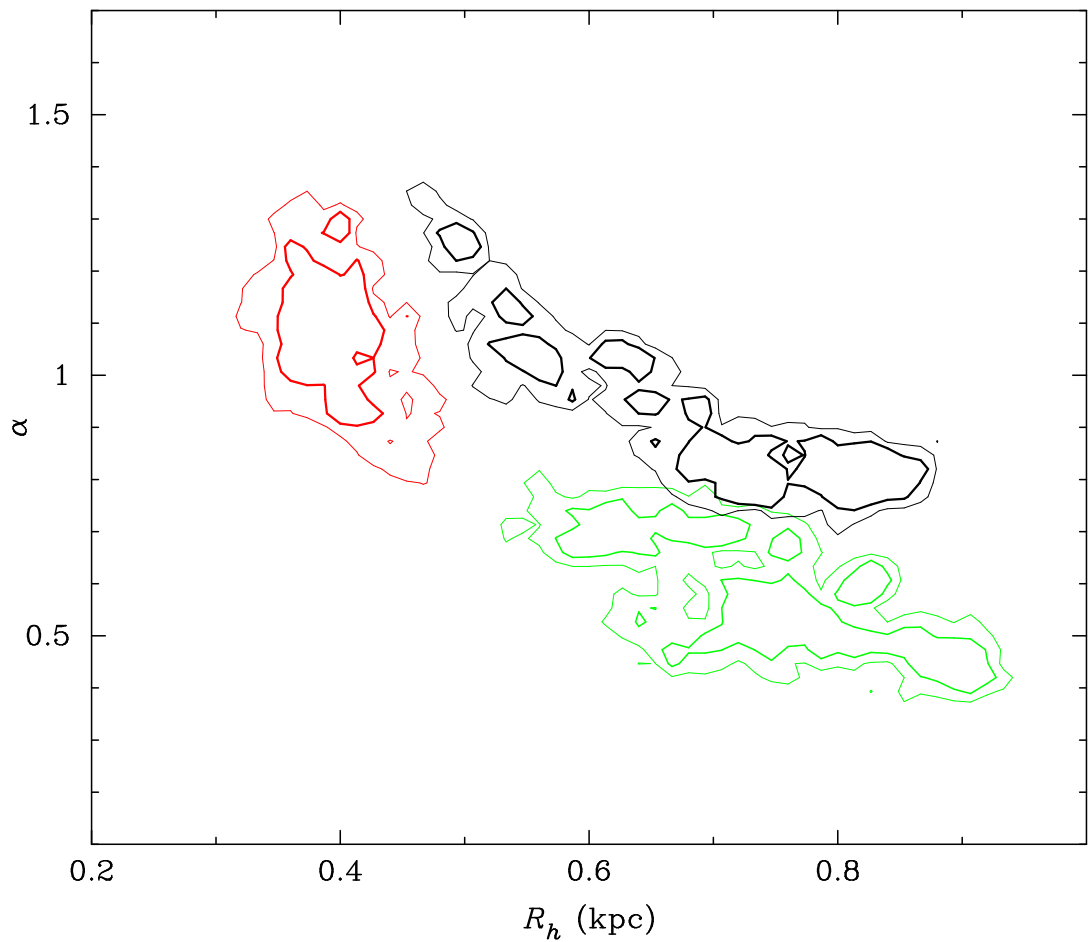}\end{center}
 \caption{The 2-D PDF of the Kormendy cutoff radius $R_h$ versus the cutoff index $\alpha$. Colours are as in Fig.~\ref{fig:kormendyall-discmass-Q}.}
 \label{fig:kormendyall-alpha-kindex}
 \end{figure}

The posterior PDFs for the  hole radius $R_h$ and the Kormendy index $\alpha$ are found in Fig.~\ref{fig:kormendyall-alpha-kindex}. We find $R_h=0.7\pm0.1$ kpc and $\alpha=0.9\pm0.1$. 
Our value for $\alpha$ is in conflict with Kormendy's (1977) result that $\alpha = 3$ for most galaxies (Baggett et al. 1998 also use $\alpha=3$ for their fits). A value of $\alpha=3$ yields a strong hole and therefore a very large bulge would be needed reproduce the SB profile. However, a large bulge would likely make the dispersion profile fit worse because the bulge is not likely to extend farther out than the highest dispersions at $\sim300$ pc. 
We note that all three runs suggest $\alpha\ll 3$; scenarios KL and KG yield $\alpha=0.6$ and $\alpha=1.1$, respectively.

\subsection{The bulge}

\begin{figure}
\begin{center}
 \includegraphics[angle=270,scale=0.8]{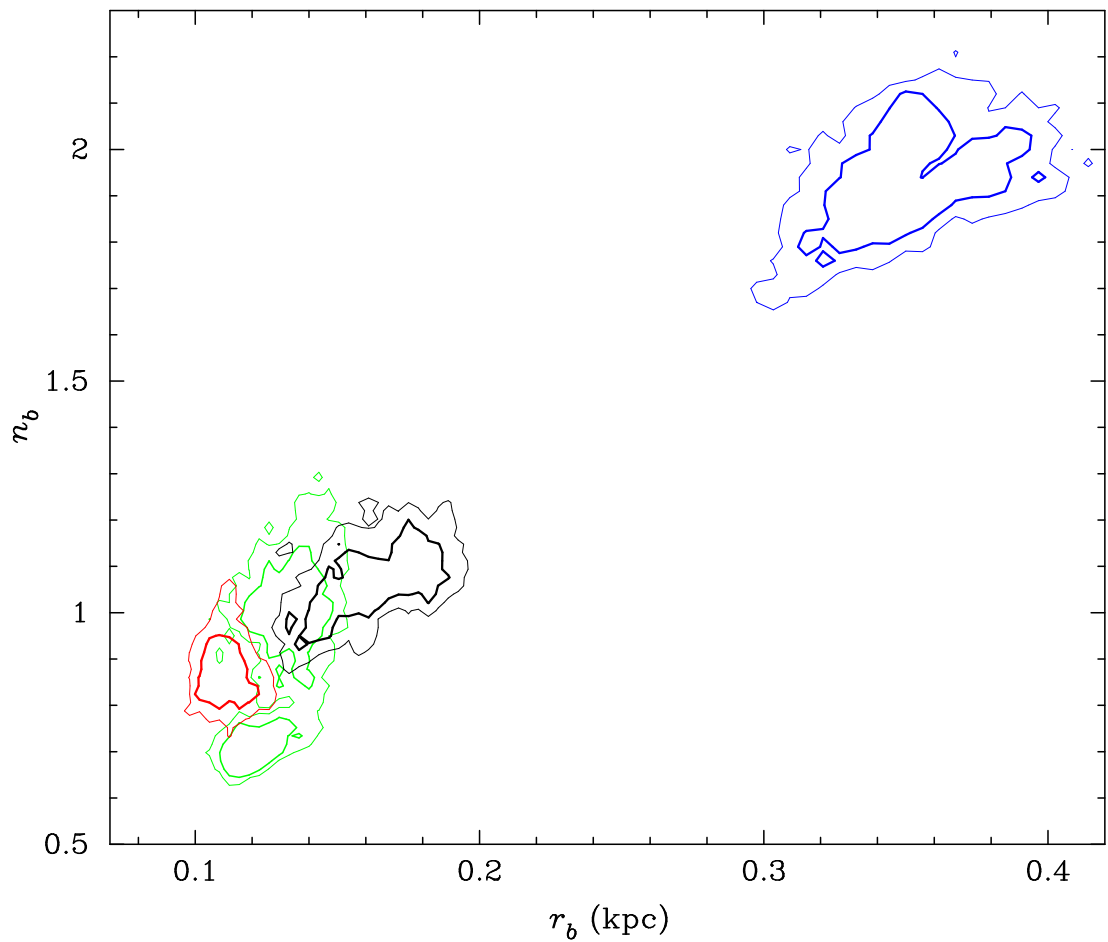}\end{center}
 \caption{The 2-D PDF of the S\'{e}rsic index $n_b$ versus the radius of the bulge $r_b$. Colours are as in Fig.~\ref{fig:kormendyall-discmass-Q}.}
 \label{fig:kormendyall-index-rb}
 \end{figure}

The results for the bulge parameters are found in Fig.~\ref{fig:kormendyall-index-rb}. For scenario K, we find a bulge radius $r_b$ of $0.16$ kpc and a S\'{e}rsic index $n_b$ of $1.1$, while scenario E yields a larger bulge with $r_b=0.35$ kpc and $n_b=1.9$. For scenarios KL and KG, the S\'{e}rsic index is $\sim0.9$. Scenarios K, KL and KG thus suggest that the bulge of NGC 6503 is nearly exponential. Note that the linear fit $n_b$ given in \S2.1 is consistent with a pure exponential bulge. 
Scenario E suggests a cuspier bulge, but even here $n_b\simeq2$, within the range of so-called `pseudobulges' \citep{kormendykennicutt}. Bulges of this type are thought to be generated via internal secular evolution processes -- the gradual buildup of gas in the centre of the galaxy -- rather than violent mergers (see, e.g., Kormendy et al. 2006). Given the NGC 6503 is isolated, secular evolution is likely the dominant evolutionary mechanism. These bulges typically resemble small discs embedded in larger discs -- their morphologies resemble flattened spheroids more than elliptical galaxies and their kinematics are rotationally dominated, but we cannot constrain the rotation of NGC 6503's bulge.

By way of comparison with known properties of bulges in spiral galaxies, the bulge radius is actually larger than found by studies of bulge-to-disc scale length correlation by roughly a factor of 2 \citep{courteauetal}. However, NGC 6503's bulge also displays certain characteristics typical of nuclear clusters, even though its size exceeds typical nuclear clusters by about two orders of magnitude \citep{walcheretal1}. Unlike normal bulges, nuclear clusters are often offset from the dynamical centre of the galaxy \citep{matthewsgallagher}. NGC 6503's bulge displays this phenomenon -- the photometric centre is offset by $\sim100$ pc. Nuclear bulges also have very small mass, typically $\lesssim5\times10^7 M_{\odot}$ \citep{walcheretal1}; the mass of NGC 6503's bulge is only slightly larger. Thus, NGC 6503 has a surprisingly low density bulge. Furthermore, \citet{walcheretal1} find that their nuclear clusters' velocity dispersions are similar to that found in NGC 6503. Their nuclear $M/L_I$ ratios are also consistent with the $M/L_R$ values we find (\S6.4), in that we expect the $M/L_I$ values to be slightly lower than our $M/L_R$ values, which is precisely what occurs. Thus, NGC 6503's bulge properties are consistent with both nuclear clusters and ordinary pseudobulges. 

The origin of nuclear bulges is unclear, but it is plausible that they are formed by secular evolution. \citet{walcheretal2} find that repeated episodes of gas infall may contribute to nuclear cluster star formation, which is required to maintain their high luminosity. In the case of scenario E, bar formation may induce gas inflow that would generate a nuclear cluster that could in turn generate the $\sigma-$drop. 
The most likely hypothesis is that of a pseudobulge that has a luminous nuclear cluster at the centre dominating the observed light.



\subsection{Halo cusps}

\begin{figure}
\begin{center}
 \includegraphics[angle=270,scale=0.8]{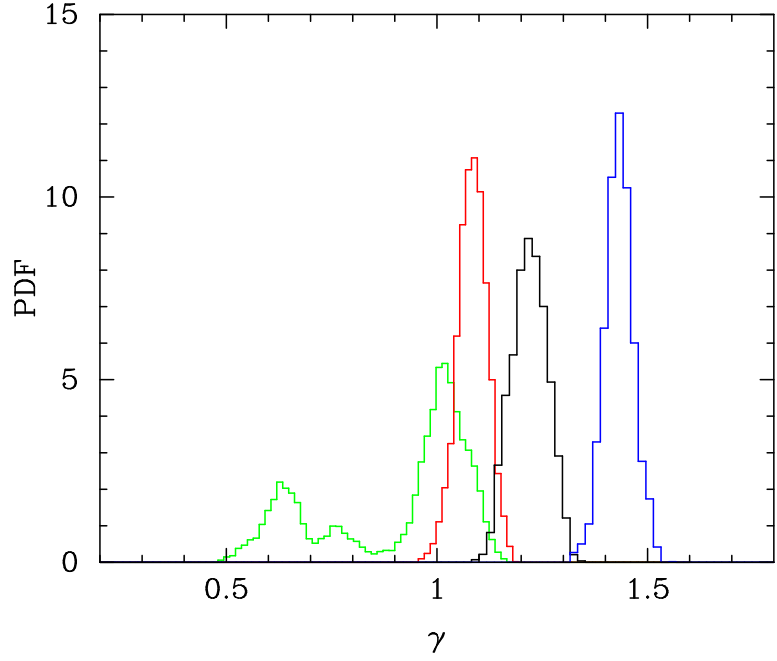}\end{center}
 \caption{The PDF for the halo cusp strength $\gamma$. Colours are as in Fig.~\ref{fig:kormendyall-heightPDF}.}
 \label{fig:kormendyall-cuspPDF}
 \end{figure}

The halo cusp $\gamma$ is well-constrained, as Fig.~\ref{fig:kormendyall-cuspPDF} attests. All runs suggest that the halo is cuspy, a result that is therefore quite robust. In the case of runs K and KL, 
we find $\gamma=1.2$ and $\gamma=0.9$, respectively. 
For scenario KG, we find that $\gamma$ is close to 1 and therefore consistent with a NFW profile; this cusp is flatter than for the equivalent scenario that fits the stellar rotation curve. 
Meanwhile, scenario E suggest a very strongly cusped halo with $\gamma=1.4$. Thus, the derived cusp value depends on the density profile of the disc and the choice of rotation curve. Unsurprisingly, a stronger cusp is required to compensate for the lack of central mass in an inner truncated disc. In addition, we find a statistically significant difference in cusp values depending on which rotation curve is fit, but both cases are consistent with or cuspier than cosmological simulations. If the assumption that gas traces the circular velocity is erroneous, then the slope of the circular velocity curve must be steeper at the centre (as seen in Fig.~\ref{fig:kormendy-model}), and hence the derived cusp value will be larger. Thus, modelling asymmetric drift correctly is essential to obtaining the correct halo density profile; we discuss possible issues with the asymmetric drift in \S7.3.

\begin{figure}
\begin{center}
 \includegraphics[angle=270,scale=0.8]{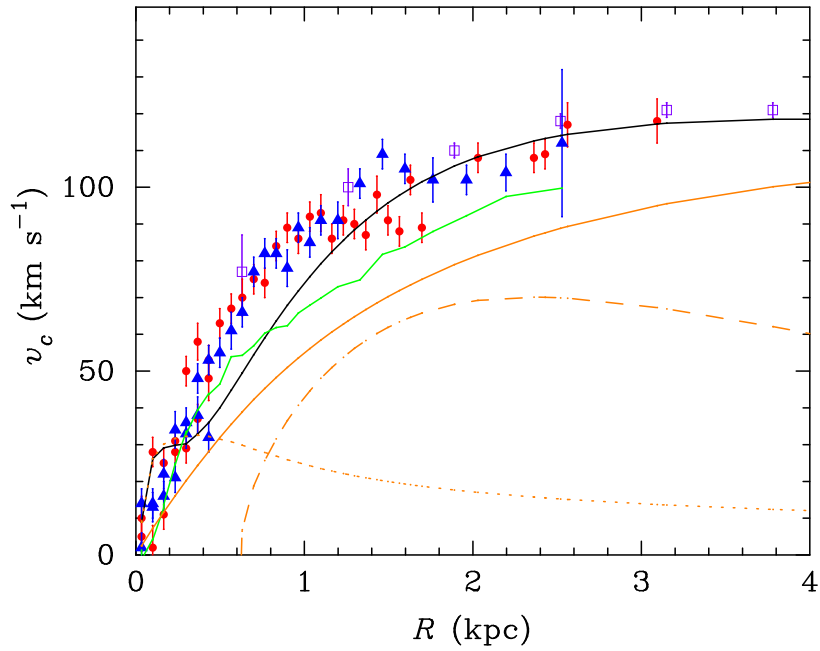}\end{center}
 \caption{Rotation curve for a model from a MCMC run in which the cusp is fixed to 0. Colours are as in Fig.~\ref{fig:kormendy-model}.}
 \label{fig:halocusp0}
 \end{figure}

To verify that the halo is cuspy, we conducted a MCMC run in which the cusp value was fixed to 0. The resulting rotation curve fit is shown in Fig.~\ref{fig:halocusp0}; the fit is clearly very poor. There is a pronounced kink in the rotation curve that is not observed, and neither the gas nor the stellar rotation curves are properly fit. We therefore reject models with $\gamma=0$.

The issue of halo cuspiness remains unsettled. It is often argued that galaxies demonstrate evidence of cored halos \citep{blaisouelletteetal, debloketal}, but many complications present themselves. Some researchers find that the rotation curves used to probe halo central structure are consistent with both cored and cuspy profiles, in part because of the size of the uncertainties \citep{vandenboschswaters, swatersetal, duttonetal, simonetal, spekkensetal}. Further caveats concern the correct interpretation of the rotation curve; the presence of noncircular gas motions, as might be caused by a triaxial halo, and the presence of turbulence caused by supernova feedback and the magnetorotational instability point to the complexity of the problem \citep{hayashietal, rheeetal, valenzuelaetal, tamburroetal}. Some steps towards resolving these issues are taken by Oh et al. 2008, who find that the non-circular motions in their sample cannot rectify the discrepancy. In fact \citet{begeman87} finds that noncircular motions in \ion{H}{1} are small in NGC 6503. However, \citet{tamburroetal} find relatively consistent values of $\sim5-25$ km s$^{-1}$ for the \ion{H}{1} dispersion across their sample of late type galaxies, and the calculated star formation for NGC 6503 (\S5.1) is consistent with their measurements. Therefore it is safe to assume that the \ion{H}{1} dispersion of NGC 6503 is $\sim 5-25$ km s$^{-1}$ and probably driven by supernova-induced turbulence inside the optical radius. Moreover, as noted in \S2.2, the \ion{H}{1} rotation is systematically larger than the H$\beta$ rotation, which suggests that the H$\beta$ rotation is more affected by these issues.

Recent simulations suggest that halo cusps are slightly shallower than previously inferred, with $\gamma=0.9\pm0.1$ \citep{navarroetal}. 
However, halos are subject to multiple effects due to baryons that are not accounted for in cosmological simulations, which may help explain why our cusp values are larger than found by Navarro et al. 
The full effect of baryonic physics on halo profiles is still poorly understood, but \citet{abadietal} find that the central density increases when more detailed baryonic effects are included. The central density cusp may also be affected by bar formation; \citet{weinbergkatz1} argue that bars are capable of washing out central cusps, an effect seen in \citet{holleybockelmannetal} and \citet{weinbergkatz2}. However, Sellwood (2003, 2008) argues that bar formation can draw mass inward, increasing the strength of the cusp; \citet{dubinskietal} find with their simulations that a live bar maintains the cusp. The differences in cusp evolution may be due to differences in the codes used. If NGC 6503 did once possess a bar, our results argue against a bar-induced cusp flattening in this galaxy.

\subsection{Masses and $M/L$ ratios}

The halo mass we find is necessarily limited by the outermost \ion{H}{1} data point at $800''$, and so should be regarded as a lower limit. The outer rotation curve is the primary determinant of the halo mass, so we do not expect significant variations in halo masses across the different runs. We find $M_{20}\sim60\times10^{9}M_{\odot}$ across all scenarios. For scenario K we obtain a $M_d=3.2\pm0.2\times10^{9}M_{\odot}$ and $M_b=7.0\pm1.4\times10^{7}M_{\odot}$. 
Larger disc masses are obtained for scenario KL, while the bulge mass is much lower for scenario KG. In scenario E, we find $M_d=3.0\pm0.4\times10^{9}M_{\odot}$ and $M_b=15\pm2\times10^{7}M_{\odot}$. 
Unsurprisingly, in all cases the halo accounts for $> 90\% $ of all the galaxy mass, while the bulge mass is too small to significantly impact the stability of the disc, except in scenario E. Whether or not the bulge masses are consistent with other values found in the literature depends on whether the bulge is interpreted as a nuclear cluster (\S6.2).

\begin{figure}
\begin{center}
 \includegraphics[angle=270,scale=0.8]{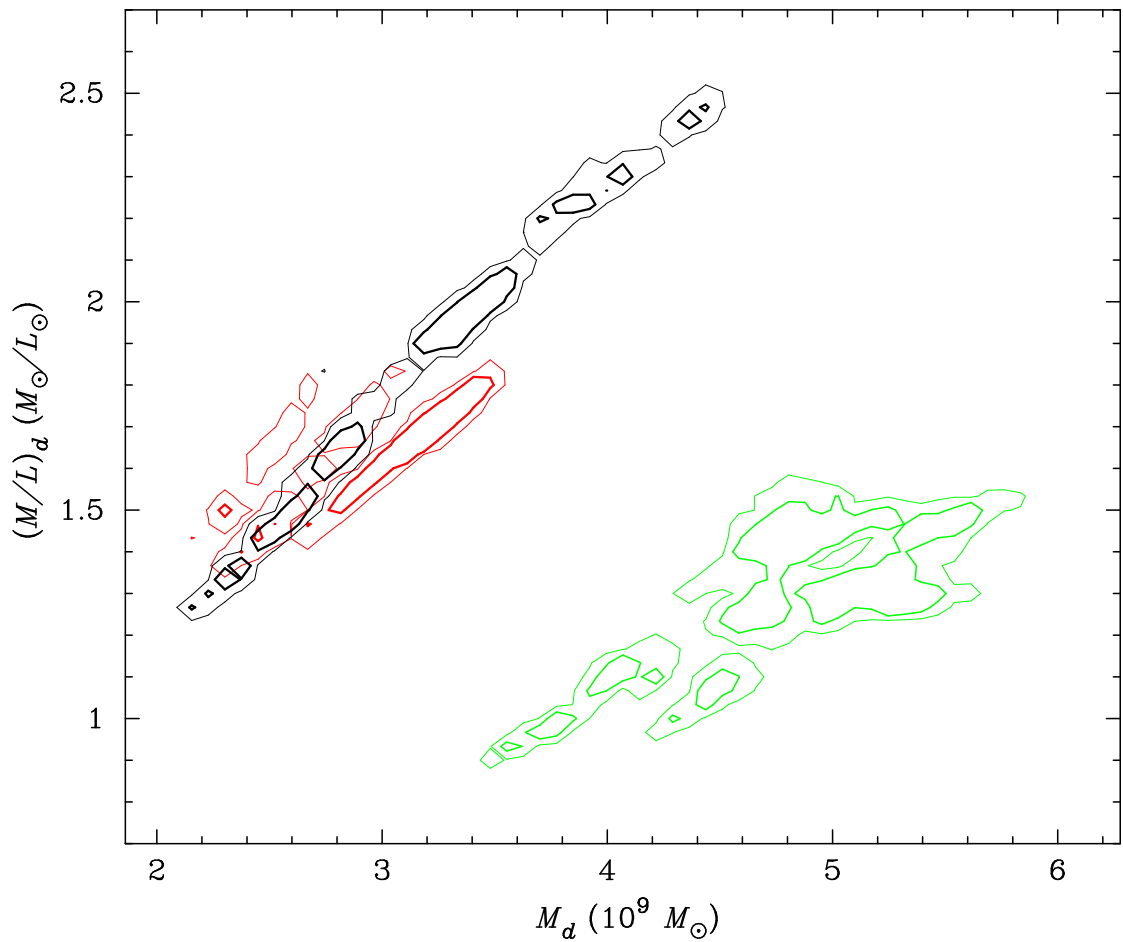}\end{center}
 \caption{The 2-D PDF of the disc mass $M_d$ versus the disc mass-to-light ratio $(M/L)_d$. Colours are as in Fig.~\ref{fig:kormendyall-discmass-Q}.}
 \label{fig:kormendyall-discmass-MLdisc}
 \end{figure}

\begin{figure}
\begin{center}
 \includegraphics[angle=270,scale=0.8]{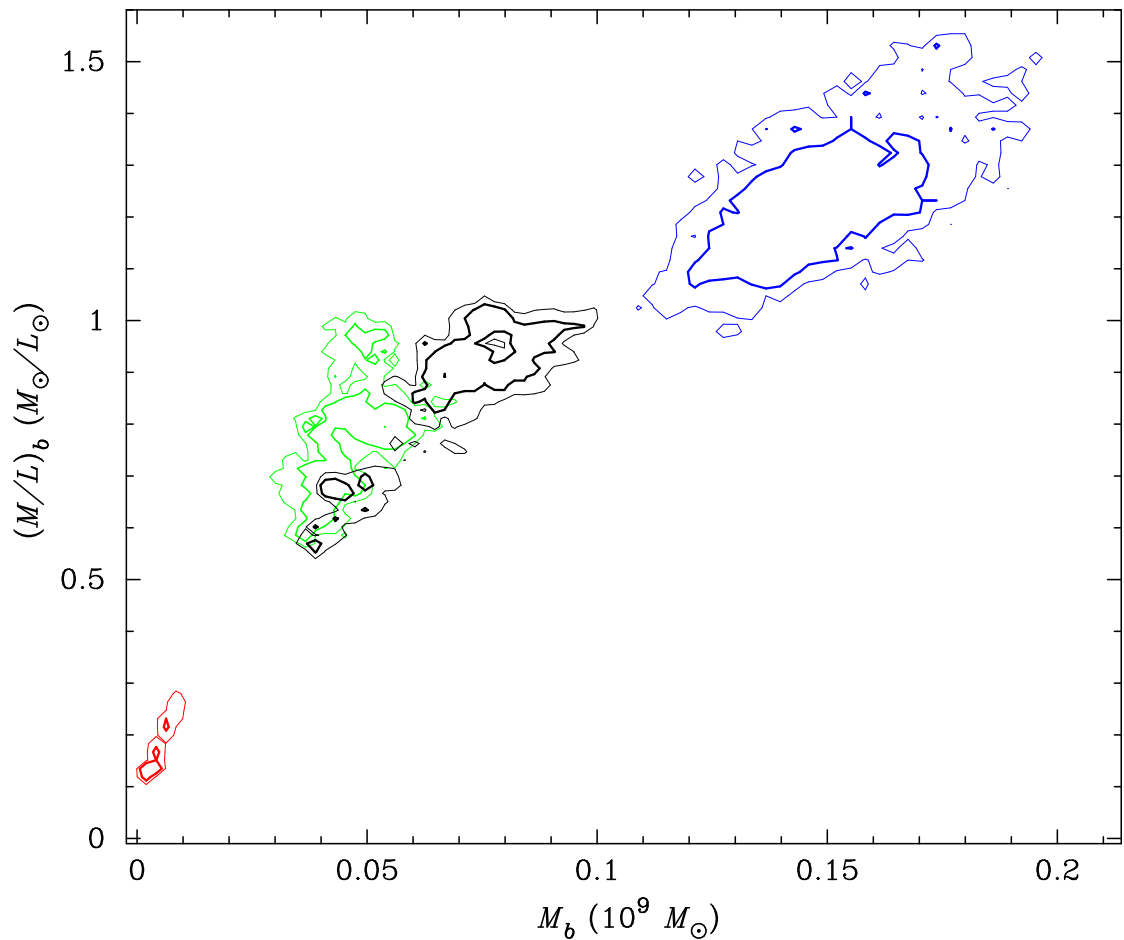}\end{center}
 \caption{The 2-D PDF of the bulge mass $M_b$ versus the bulge mass-to-light ratio $(M/L)_b$. Colours are as in Fig.~\ref{fig:kormendyall-discmass-Q}.}
 \label{fig:kormendyall-bulgemass-MLbulge}
 \end{figure}

$M/L$ ratios for the disc and bulge are shown in Figs.~\ref{fig:kormendyall-discmass-MLdisc} and \ref{fig:kormendyall-bulgemass-MLbulge}. In order to reproduce the $\sigma-$drop, the bulge $M/L$ ratios must be small, as we find in scenarios K, KL and KG, but not E. The first three scenarios show that the bulge $M/L$ is less than 1, while the disc $M/L$ is roughly twice that (for scenario KG, it is an order of magnitude larger). We find the disc $M/L_R$ to be $1.9\pm0.3 $ and the bulge $M/L_R$ to be $0.9\pm0.1$ under scenario K. In scenario E, we fixed the disc $M/L_R$ to be $0.53M_d$, yielding a disc $M/L_R$ of $1.6\pm0.2$ while the bulge $M/L_R$ is $1.2\pm0.1$, which, although larger than 1, is still lower than the disc $M/L_R$. 

Our models assume constant $M/L$ throughout the galaxy disc in the absence of a stellar population gradient. In particular, stellar $M/L$ ratios tend to increase in spiral galaxy centres where redder colours prevail \citep{belldejong, dejongbell}. From B89, $B-R$ across two disc scale lengths for NGC 6503 varies from 1.10 to 1.41, correlating with $M/L_R$ ratios of $\sim 1.3-2.7$ using Table 3 of \citet{belldejong}, which is consistent with our findings. 
Upper limits to the dynamical $M/L_R$ values were obtained by \citet{broeilscourteau}, who find $M/L_R\lesssim 4$, also consistent with our values.

The $M/L$ values for the bulge do not, by contrast, coincide with other bulge values found in the literature for nearby spiral galaxies. \citet{yoshinoichikawa} generally find that their bulge $M/L$ ratios are larger than their disc values with few exceptions; $M/L_V$ ratios are $\sim 4.5\pm 2.4$ and $M/L_I$ ratios are $\sim 2.7\pm 1.8$. In no case is the bulge $M/L_V$ value lower than 1, and only in a few cases does $M/L_I$ fall below 1. NGC 6503's bulge $M/L$ lies significantly below nearly all the bulges in this sample. However, $M/L$ ratios are highly sensitive to the formation of massive, luminous stars, and our small $M/L$ values, together with the work of \citet{wozniaketal} and \citet{champavertwozniak} showing that $\sigma-$drops can be caused by massive star formation, strongly suggest that the bulge of NGC 6503 has a star-forming component. Thus, the better comparison may be to the nuclear clusters of \citet{walcheretal1}, who find $M/L_I$ ratios that generally lie below 1. 

\section{Discussion}

\subsection{Comparison with earlier work}

B89 and BG97 present dynamical models for NGC 6503, and BG97 test the models for bar stability. In both papers, the mass distribution of the disc is given by Eq.~\ref{eq:expsech} with a scalelength of $R_d=1$ kpc, while the velocity ellipsoid is given by
\begin{eqnarray}
\label{eq:iso}
\sigma_z &=& \sqrt{\pi \Sigma(R) z_d}  \\
\label{eq:iso2}
\sigma_R &=& \sigma_z / 0.6 \\
\sigma_{\phi} &=& \sigma_R \sqrt{B/(B-A)}
\end{eqnarray}
where $\Sigma(R)$ is the projected surface density, $z_d$ is the scale height of the disc, and $A$ and $B$ are the Oort constants \citep{vanderkruitsearle}. The expression for $\sigma_z$ is obtained for an isolated isothermal disc, while the relationship between $\sigma_R$ and $\sigma_{\phi}$ comes from the epicycle equations. However, the relationship between $\sigma_R$ and $\sigma_z$ is an extrapolation from the observed value in our solar neighbourhood. We return to this assumption in \S7.2.

B89 also examined a second model in which $Q$ is held constant throughout the disc \citep{carlbergsellwood}. For this model, the radial dispersion obeys
\begin{equation}
 \sigma_R \propto \frac{\rho}{\kappa}\exp(-R/R_d)\left(B(B-A)\right)^{-1/2}. 
\end{equation}
Both models yield reasonable fits (the former is slightly better), but cannot reproduce the $\sigma-$drop. To reproduce a borderline stable disc where $Q=1.7$ \citep{sellwoodcarlberg}, B89 finds that the disc $M/L$ must be $1.7\pm0.3$ in the $B-$band, which is consistent with our findings.

BG97 model the galaxy in more detail and provide stability studies. They adopt the gas rotation curve as the fundamental input to their simulations and assume an isothermal, cored halo profile along with the disc from B89. They vary the disc to halo mass ratio to investigate the stability properties of the galaxy; $Q$ declines slightly from the lowest disc to halo mass ratio to the highest.

The dispersions of the isolated disc scale with the square root of its surface density (Eqs.~\ref{eq:iso} and \ref{eq:iso2}), but the process of embedding the disc in a dark halo modifies the dispersions. BG97 find that embedding the isolated disc in the halo allowed the settling process at the beginning of the $N-$body simulations to rearrange the dispersions itself.\footnote{BG97 attempted to account for the embedding: $\sigma_z$ can be modified by a factor $F(\epsilon)$ accounting for the relative halo contribution, where 
\begin{equation}
\epsilon = \left(\rho_{\mathrm{halo}}^{z=0} - \frac{1}{4\pi R}\frac{\partial (v_C^2)}{\partial R}\right)/\left(\rho_{\mathrm{disc}}^{z=0}\right).
\end{equation}
The result is that when $\epsilon$ is less than 0 (as it could be near the centre where $\frac{\partial}{\partial R}v_c^2$ is very large) the dispersion can decrease upon embedding, while large values of $\epsilon$ cause the dispersions to increase. 
BG97 dropped this factor for the models discussed in that paper and used the isolated disc dispersions (Bottema, private communication).} The result is an initial central outflow in the simulations, leading to a decrease in the dispersions, more-or-less naturally accounting for the observed disc dispersion of NGC 6503. BG97 find that their two lightest discs remain stable to bars, but are unable to reproduce the $\sigma-$drop.

Visible on Fig.~\ref{fig:kormendyall-discmass-Q} are the points corresponding to the models used by BG97. They are located well above the $2\sigma$ confidence interval. The discrepancy occurs because $Q$ is obtained from the radial dispersion $\sigma_R$, which, in BG97's models, is directly coupled to the surface density (cf. Eqs.~\ref{eq:iso} and \ref{eq:iso2}). Thus, BG97's $\sigma_R$ is determined by the disc surface density, which is not the case for the GalactICS model. 
Their Fig.~9 shows that, after settling, 
their minimum $Q-$values decline so that they fall in line with our PDF in Fig.~\ref{fig:kormendyall-discmass-Q}. Thus, their $M_d-Q_{\mathrm{min}}$ data points are consistent with our $M_d-Q_{\mathrm{min}}$ PDF. However, their discs do not display bar formation, while ours do; there are two possible reasons for this. First, from Fig.~\ref{fig:kormendyall-bar}, most of our models only develop a bar after $\sim 2 $ Gyr, while BG97 evolved their simulations for $\sim 1.3$ Gyr. Second, and more importantly, it is now known that live halos can trigger bar formation \citep{athanassoula02}; BG97 used rigid halo potentials for their simulations, so this bar formation trigger is not in play. We evolved one of our bar unstable models in exactly the way described in \S5, except the number of disc particles was reduced to 40K. We found that the bar formation in this model was delayed by $\sim 1-2$ Gyr, suggesting that sufficient numerical resolution is needed to capture the correct evolution. This is consistent with halo-triggered bar formation, because inadequate numerical resolution will fail to resolve the resonant interactions required for this mechanism.

\subsection{The ratio of velocity dispersions}

The ratio $\sigma_z/\sigma_R=0.6$ is an observation of the solar neighbourhood. Should this ratio hold at all radii in a general disc galaxy? \citet{gerssenetal1} and \citet{gerssenetal2} find that the ratio is closer to 0.7 for NGC 488 and to 0.85 for NGC 2985, and \citet{westfalletal} find that the ratio is also high for NGC 3949 and NGC 3982 (the latter's vertical dispersion is larger than its radial dispersion). Theoretical studies of the dispersion ratio have been carried out by \citet{idaetal} and \citet{shiidsukaida}, who find that $\sigma_R/\sigma_z \sim 0.6$ is roughly correct provided that $\kappa/\Omega \lesssim 1.5$; higher values of $\kappa/\Omega$ imply higher $\sigma_z$ values. This condition is probably satisfied in the centres of galaxies, and it would therefore be unreasonable to extend a fixed ratio of velocity dispersions across the entire disc. In fact, one might surmise that the reason for the central outflow in BG97's heavier discs is the incorrect ratio of velocity dispersions there.

\begin{figure}
\begin{center}
 \includegraphics[angle=270,scale=0.8]{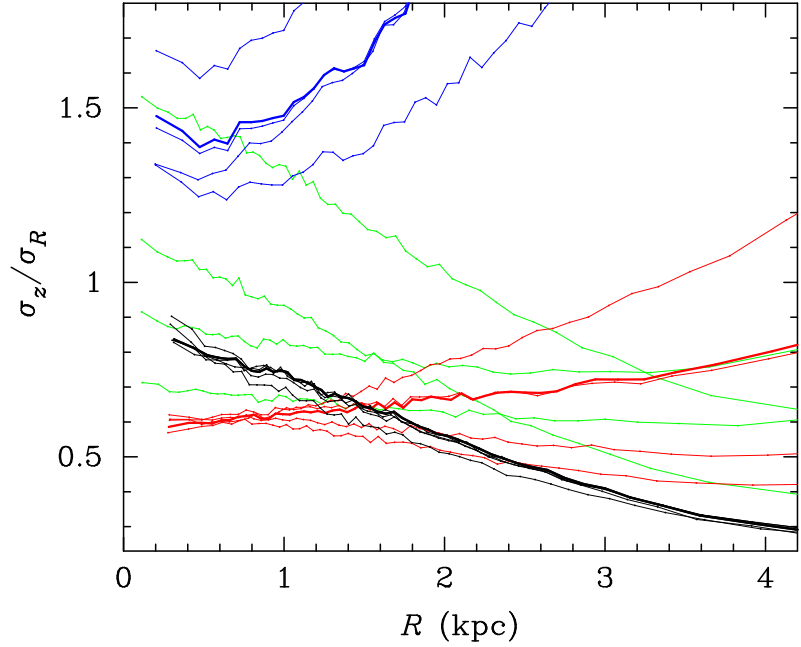}\end{center}
 \caption{Ratio of velocity dispersions for NGC 6503. Black refers to scenario K, green refers to scenario KL, red refers to scenario KG and blue refers to scenario E. The thick lines use the best fit values found in Table~\ref{table:resultsall} for scenario K and the best fit values for scenario KG (not published), while the thin lines correspond to the specific models identified by stars on Fig.~\ref{fig:kormendyall-discmass-Q}. There is no thick line for scenario KL because the means for scenario KL typically fall between two separate PDFs (\S5).}
 \label{fig:dispersionratio}
 \end{figure}

To examine the velocity dispersion ratio for NGC 6503, we took the fiducial best fit model for NGC 6503 (that is, the means from Table~\ref{table:resultsall}) and all models used to test bar stability and calculated the ratio as a function of radius. The results are found in Fig.~\ref{fig:dispersionratio}. NGC 6503's dispersion ratio depends critically on the assumed model -- scenario K exhibits a nearly linear decline with radius in which $\sigma_z/\sigma_R>0.6$ within two disc scale lengths. There is no indication of flattening that would justify adopting a single value over the entire disc, and very little scatter. Scenarios E, KL and KG display considerably more scatter, although only scenario KG displays anything close to a relatively constant ratio of 0.6 over a significant range; models from scenario KG appear to converge to $\sigma_z/\sigma_R\sim0.6$ at small radii, but the scatter increases at larger radii. The shape and slope of the curve is critically dependent on the choice of $R_d$ and $R_{\sigma}$; in scenario K, $R_d\ll R_{\sigma}$, which drags down the dispersion ratio at outer radii. If we set $R_{\sigma}$ to be 50\% of the fiducial value, the ratio would fall less rapidly with radius but would still not be constant. Note that when $R_d=R_{\sigma}$ the ratio is constant by Eqs.~\ref{eq:iso} and \ref{eq:iso2}, although this may not hold in an embedded disc. Overall, we find that a constant dispersion ratio does not apply for this galaxy.

\subsection{The relationship between asymmetric drift, stability and cusp value}

The issues of cusps, asymmetric drift and stability are intertwined because all hinge on the reliability of rotation curve data and all depend on a full accounting of effects that alter the interpretation of the rotation curve. As already indicated, it is inappropriate to assume that the gas rotation traces the gravitational potential, implying that: (1) the difference between the gas and stellar rotation curves is not a measure of the asymmetric drift; (2) the gas rotation cannot be used to assess the cusp value; and (3) the proper choice of rotation curve to model is critical to finding the correct stability region. As noted in \S2.2, the theoretical asymmetric drift (Eq.~\ref{eq:vasym}) is much larger than the difference between the gas and stellar rotation curves. We have conducted separate MCMC runs in which we fit both the gas rotation curve as a tracer of the circular velocity \emph{and} the stellar velocity. This only yields models with $Q<1$, because of the very low asymmetric drift implied by the nearly coincident rotation curves. Such models, when simulated, display sharp instabilities including disc fragmentation and cannot reproduce the observed properties of the galaxy. Our preference for fitting the stellar rotation curve and finding that gas cannot be assumed to trace to circular velocity is corroborated by \citet{pizzellaetal}. They analyse rotation curves of LSB galaxies and find that stellar rotation curves are much more regular and amenable to modelling than gas curves, which suffer from numerous issues including noncircular and vertical motions that affect their speeds relative to the circular velocity. 

The impact on the cusp value is notable because, if gas does not trace the gravitational potential, the circular velocity is steeper in the centre than the slope of the gas rotation. Assuming that the gas traces the gravitational potential will therefore lower the inferred cusp value. This is a possible source of error in early work assessing the observed cusp value in LSB galaxies, as small changes in central slope can lead to large changes in the cusp value or possibly imply a cored halo ($\gamma=0$). We avoid this problem by using a model that follows the
stellar kinematics with asymmetric drift incorporated self-consistently. 
Once the gas asymmetric drift is properly accounted for by including a full treatment of noncircular motions and turbulence, gas rotation curves should be as reliable as stellar rotation curves. Detailed measurements of the \ion{H}{1} dispersion in spiral galaxies are becoming more common and will aid the modelling of gas kinematics. For example, \citet{boomsmaetal} obtain the high resolution \ion{H}{1} velocity map for NGC 6946, finding dispersions of $\sim 6-13$ km s$^{-1}$, high velocity \ion{H}{1} clumps that lag the disc rotation, and hundreds of ``holes'' in the \ion{H}{1} distribution which are likely due to star formation. We surmise that modelling these phenomena and determining the impact on derived galaxy parameters is a nontrivial exercise. However, while easier to model than the gas asymmetric drift, the stellar asymmetric drift is also subject to several complications, a few of which we now address.

\begin{figure}
\begin{center}
 \includegraphics[angle=270,scale=0.8]{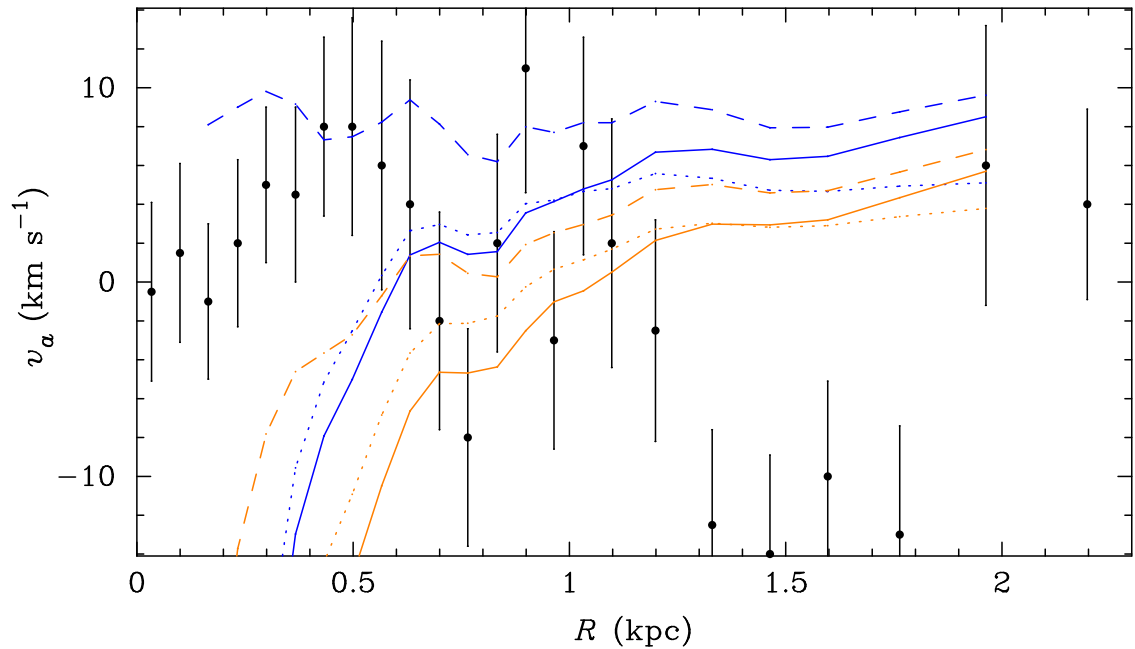}\end{center}
 \caption{Calculated asymmetric drifts in the case of cylindrical alignment and Kormendy hole (solid blue), cylindrical alignment and exponential disc (dashed blue), cylindrical alignment, Kormendy hole and low $R_{\sigma}$ (dotted blue), spherical alignment and Kormendy hole (solid orange), spherical alignment and exponential disc (dashed orange), and spherical alignment, Kormendy hole and low $R_{\sigma}$ (dotted orange),  using the fiducial parameters for scenario K found in Table~\ref{table:resultsall}. The black points are the observed asymmetric drift, given by the difference between the H$\beta$ rotation and the stellar rotation.}
 \label{fig:vasym}
 \end{figure}

Eq.~\ref{eq:vasym} is derived assuming an exponential disc, cylindrical alignment of the velocity ellipsoid and $R_d=R_{\sigma}$. With an inner truncated disc and $R_d\neq R_{\sigma}$, the asymmetric drift becomes
\begin{equation}
 v_a = \frac{\sigma_R^2}{2v_c}\left[\frac{R}{R_d}+\frac{R}{R_{\sigma}}+\frac{1}{2}\left(\frac{R}{v_s}\frac{\partial v_s}{\partial R} + 1\right) - 1 -  \alpha\left(\frac{R_h}{R}\right)^{\alpha}\right].
\label{eq:vasym1}
\end{equation}
If we now assume that the velocity ellipsoid is spherically aligned, we obtain 
\begin{equation}
 v_a = \frac{\sigma_R^2}{2v_c}\left[\frac{R}{R_d}+\frac{R}{R_{\sigma}}+\frac{1}{2}\left(\frac{R}{v_s}\frac{\partial v_s}{\partial R} + 1\right) - \alpha\left(\frac{R_h}{R}\right)^{\alpha} - 2 + \frac{\sigma_z^2}{\sigma_R^2}\right].
\label{eq:vasym2}
\end{equation}
This expression will match Eq.~\ref{eq:vasym1} only if $\sigma_z=\sigma_R$, which is not generally the case (\S7.2). Both expressions reduce the asymmetric drift relative to what Eq.~\ref{eq:vasym} predicts; in particular, the addition of an inner truncation reduces the expected drift significantly in the inner parts. The size of the change is only a few km s$^{-1}$, but it is still a significant fraction of the predicted drift. This is seen in Fig.~\ref{fig:vasym}, in which the effects of adding an inner truncation and spherical alignment of the velocity ellipsoid are examined. It is noteworthy that the inner truncation can effectively wipe out the asymmetric drift over a large range in radius; a similar effect is seen in Fig.~\ref{fig:kormendy-model}. For Fig.~\ref{fig:vasym}, we use the fiducial values for scenario K found in Table~\ref{table:resultsall}. Note that the increased asymmetric drifts at larger radii are due to the large $R_{\sigma}$ of this scenario, and the asymmetric drift expression breaks down inside $\sim0.5R_d$. We test cases in which $R_{\sigma}$ is reduced to 50\% of the fiducial value and find that the asymmetric drift is reduced in the outer part. Thus, as is evident from Fig.~\ref{fig:vasym}, a combination of factors can serve to reduce the asymmetric drift nearly to the observed values in this galaxy. One should keep in mind that the sensitivity of the asymmetric drift expression to such changes means that the predicted drifts may fluctuate wildly depending on the assumptions. 

Because of these issues, the circular velocity derived in scenarios K, KL and E may be larger than the true circular velocity. These caveats further suggest that the asymmetric drift found in scenario KG may be smaller than seen in Fig.~\ref{fig:kormendy-model}, and hence that scenario KG may be potentially viable. Without more information about the asymmetric drift in real galaxies, it is impossible to claim that scenario KG is definitely unrealistic; however, scenarios K and E remain our preferred models for this galaxy.

\subsection{Evidence for a bar}

Recent work by \citet{freelandetal} suggests that NGC 6503 may possess a nuclear disc of radius $\sim100$ pc and an end-on bar; deprojection of their $H-$band image of NGC 6503 shows a bar like structure perpendicular to the major axis with clear spiral arms emanating from the ends. This scenario is consistent with our scenario E. It would moreover explain the $\sigma-$drop because of star formation at the centre caused by bar induced gas inflow and because diagnostics from \citet{bureauathanassoula} suggest that end-on bars may generate $\sigma-$drops. These findings are compelling evidence for a bar in NGC 6503. Notably, these authors also estimate the age of the star formation ring at $\sim 0.5$ Gyr, which is well below the timescale for bar destruction determined by \citet{bournaudetal} of $\sim1$ Gyr.

The presence of a bar in NGC 6503 is unlikely to change our main conclusions because it is end on. There is no clear kinematic indicator of bar structure in the stellar or ionized gas rotation curve; we expect that the measured values are largely unaffected by the bar since only a small region of the slit used for observation would have intersected the bar. Part of the $\sigma-$drop could be due to the bar, and thus the bulge
may not be as large and have as low a density as found in \S6.2. Other bulge parameters are less likely to be affected by the bar; in particular, the conclusion that there is a star forming component in the galaxy centre is unchanged.

\section{Conclusion}

We have deployed a Bayesian/Markov chain Monte Carlo technique to model the disc galaxy NGC 6503. We find models for NGC 6503 that satisfy the observational constraints using this technique, and are able to constrain the input parameters. The data includes a Freeman Type II surface brightness profile, ionized gas and \ion{H}{1} rotation curves, the stellar rotation curve and the stellar line of sight velocity dispersion. Four different scenarios were considered: an inner truncated disc, an exponential disc with dust, a model in which gas traces the gravitational potential, and a model in which the true underlying disc scale length is revealed in the outermost portion of the surface brightness profile. We find that the second scenario leads to models that strongly bar unstable, and the third scenario cannot reproduce the stellar rotation curve. The first scenario provides the best fit to the data and is most likely to offer the bar stability required to correctly model the galaxy, but the last scenario also provides a realistic model for the galaxy. 

Further properties of the galaxy are discerned; the bulge is a pseudobulge, and the bulge $M/L$ is lower than the disc $M/L$, suggesting a star forming component that is probably responsible for the $\sigma-$drop. We also find that the halo must be cusped with $\gamma\gtrsim1$, a result that is robust to all fitting methods. The Bayesian/MCMC technique used to discover these results is robust and flexible; we find it to be an effective tool for fitting galaxy models in complex parameter spaces.

\acknowledgements

We thank K. Spekkens, P. Teuben, R. Bottema, J. Dubinski, M. Bershady and C. Arsenault for useful and informative discussions. S.C. and L.M.W. wish to acknowledge support through respective Discovery Grants from the Natural Sciences and Engineering Research Council of Canada.



 
\clearpage

\begin{table}
\begin{center}
\begin{tabular}{lll}
 \tableline
Parameter&Units&Description\\
\tableline
$r_h$&kpc&Halo truncation radius\\
$v_h$&100 km s$^{-1}$&Halo characteristic velocity\\
$a_h$&kpc&Halo characteristic radius\\
$\delta$&kpc&Halo truncation width\\
$\gamma$&Dimensionless&Cusp value\\
$M_e$&$10^9 M_{\odot}$&Exponential disc mass\\
$R_d$&kpc&Photometric disc scale length\\
$z_d$&kpc&Disc scale height\\
$n_b$&Dimensionless&S\'{e}rsic index\\
$v_b$&100 km s$^{-1}$&Bulge characteristic velocity\\
$r_b$&kpc&Bulge characteristic radius\\
$\sigma_0$&100 km s$^{-1}$&Central velocity dispersion\\
$b_{\mathrm{rot}}$&Dimensionless&Bulge rotation parameter\\
$(M/L)_d$&$M_{\odot}/L_{\odot}$&Disc mass to light ratio\\
$(M/L)_b$&$M_{\odot}/L_{\odot}$&Bulge mass to light ratio\\
$R_\sigma$&kpc&Radial dispersion disc scale length\\
$R_h$&kpc&Kormendy cutoff radius\\
$\alpha$&Dimensionless&Kormendy cutoff index\\
\tableline
Calculated Quantities\\
\tableline
$M_d$&$10^9 M_{\odot}$&Real disc mass\\
$M_b$&$10^9 M_{\odot}$&Model bulge mass\\
$M_{20}$&$10^9 M_{\odot}$&Model halo mass within 20 kpc\\
$Q$&Dimensionless&Toomre local stability parameter\\
$X$&Dimensionless&Toomre global stability parameter\\
\tableline
\end{tabular}
\end{center}
\caption{Table of the input parameters for the MCMC chains, as well as calculated quantities that are output for each model.}
\label{table:parametertable}
\end{table}

\begin{table}
\begin{center}
\begin{tabular}{lrlrlrl}
\tableline
Parameter & Best-fit value &  &  &  & Observed value &  \\ 
\tableline
 & Scenario K &  & Scenario E & \\
\tableline
$v_h$ & 2.47 & $\pm0.02$ & 2.38 & $\pm0.02$ &  & \\
$a_h$ & 6.66 & $\pm0.49$ & 8.69 & $\pm0.66$ &  &\\
$\gamma$ & 1.22 & $\pm0.04$  & 1.43 & $\pm0.03$ &  & \\
$M_e$ & 5.77 & $\pm0.83$  & 3.12 & $\pm0.43$ &  & \\
$R_d$ & 0.71 & $\pm0.02$  & 1.3 & \textit{~fixed} & 1 &\\
$z_d$ & 0.14 & $\pm0.01$  & 0.24 & $\pm0.02$ & $<0.2R_d$ &\\
$n_b$ & 1.06 & $\pm0.07$  & 1.93 & $\pm0.10$ &  &\\
$v_b$ & 0.43 & $\pm0.04$  & 0.51 & $\pm0.03$ &  &\\
$r_b$ & 0.16 & $\pm0.02$  & 0.35 & $\pm0.02$ &  &\\
$\sigma_0$ & 0.37 & $\pm0.03$   & 0.34 & $\pm0.03$ & 0.55 & $\pm0.10$\\
$(M/L)_d$ & 1.86 & $\pm0.33$   & 1.60 & $\pm0.2$ & 1.41 & $\pm$factor of 2 \\
$(M/L)_b$ & 0.87 & $\pm0.12$   & 1.23 & $\pm0.11$ &  &\\
$R_{\sigma}$ & 2.57 & $\pm0.29$   & 0.75 & $\pm0.05$ & 1 &\\
$R_h$ & 0.69 & $\pm0.11$   &  &  &  &\\
$\alpha$ & 0.90 & $\pm0.14$   &  &  &  &\\
\tableline
$M_d$& 3.19 & $\pm0.59$ & 3.03 & $\pm0.39$ &  &\\
$M_b$& 0.070 & $\pm0.014$ & 0.15 & $\pm0.02$ &  &\\
$M_{20}$& 61 & $\pm2$ & 61 & $\pm2$ &  &\\
$Q$ & 1.62 & $\pm0.24$ & 0.69 & $\pm0.15$ & 1.3 &\\
$X$ & 1.88 & $\pm0.46$ & 2.84 & $\pm0.57$ &  & \\
\tableline
\end{tabular}
\end{center}
\caption{The final posterior values for each input and calculated parameter with $1\sigma$ error bars, with units as in Table~\ref{table:parametertable}. Where possible, the observed values are obtained from B89 or (in the case of the disc mass to light ratio) BG97.}
\label{table:resultsall}
\end{table}

\end{document}